\documentclass[journal]{IEEEtran}

\usepackage{graphicx,latexsym,color}
\usepackage{amsmath,amsthm,amsfonts}
\usepackage{latexsym}
\usepackage{cite,citesort}
\usepackage[normalem]{ulem}	% Part of the standard distribution

\definecolor{MyDarkBlue}{rgb}{0.1,0,0.55} 
\definecolor{MyRed}{rgb}{1,0,0} 
\definecolor{MyBlue}{rgb}{0,0,1} 
\definecolor{MyGreen}{rgb}{0,1,0} 
\usepackage{bm}
\usepackage{tikz}
\usepackage{pgfplots}
\usetikzlibrary{mindmap,trees}
\usetikzlibrary{%
  3d,
  calc,
  arrows,%
  shapes.misc,% wg. rounded rectangle
  shapes.arrows,%
  chains,%
  matrix,%
  positioning,% wg. " of "
  scopes,%
  decorations.pathmorphing,% /pgf/decoration/random steps | erste Graphik
  shadows%
}

\renewcommand{\vec}[1]{{\boldsymbol#1}}

\newcommand{\ie}{\textit{i.e.}\/, }
\newcommand{\eg}{\textit{e.g.}\/, }
\newcommand{\cf}{\textit{cf.}\/, }

\providecommand*{\mrm}[1]{\mathrm{#1}}
\providecommand*{\unit}[1]{\ensuremath{\mrm{\,#1}}}
\providecommand*{\eu}{\ensuremath{\mrm{e}}}
\providecommand*{\iu}{\ensuremath{\mrm{i}}}

\providecommand*{\diff}{\operatorname{d}\!}
\renewcommand{\Im}{\operatorname{Im}}	% The LaTeX standard is not ISO!

\providecommand*{\degree}{\ensuremath{^\circ}}

\begin{document}

\title{A quasi-static electromagnetic analysis for experiments with strong permanent magnets}

\author{Sven~Nordebo,~\IEEEmembership{Senior Member,~IEEE}, 
Alexander~Gustafsson
\thanks{Manuscript received \today. %The authors acknowledge the comments and suggestions provided by Patrik Wahlberg.
} 
\thanks{Sven~Nordebo and Alexander~Gustafsson are with the Department of Physics and Electrical Engineering, Linn\ae us University, 
351 95 V\"{a}xj\"{o}, Sweden. (E-mail: \{sven.nordebo,alexander.gustafsson\}@lnu.se).}
}

%\markboth{Submitted to IEEE Transactions on Education  \today.}{}

\maketitle

\begin{abstract}
An electromagnetic analysis is presented for experiments with strong permanent disc magnets.
The analysis is based on the well known experiment that demonstrates the effect of circulating eddy currents
by dropping a strong magnet through a vertically placed metal cylinder
and observing how the magnet is slowly falling through the cylinder with a constant velocity.
This experiment is quite spectacular with a super strong neodymium magnet and a thick metal cylinder
made of copper or aluminum.
A rigorous theory for this experiment is provided based on the quasi-static approximation
of the Maxwell equations, an infinitely long cylinder (no edge effects) and a homogeneous magnetization of the disc magnet.
The results are useful for teachers and students in electromagnetics who wish to 
obtain a deeper insight into the analysis and experiments regarding this phenomenon, or with industrial applications
such as the grading and calibration of strong permanent magnets or with measurements of the conductivity of various metals, etc.
Several experiments and numerical computations are included to validate and to illustrate the theory.
\end{abstract}


\section{Introduction}\label{sect:introduction}
\IEEEPARstart{A}{} well known physics/electromagnetics experiment to 
demonstrate the effect of circulating eddy currents
is to drop a strong magnet through a vertically placed metal cylinder
and to observe how the magnet, quite unexpectedly, is not accelerating as usual
but is only slowly falling through the cylinder with a constant velocity.
The stronger the magnet, or the lower the resistance of the metal cylinder, the slower is the fall through the cylinder.
The experiment demonstrates that circulating eddy currents are induced inside the metal cylinder due to the
changing magnetic flux, \ie the Faraday's law of induction, and that these induced currents cause a secondary 
magnetic field and associated magnetic forces that oppose the fall of the magnet, all in accordance to Lenz's law
as well as the Biot-Savart law  \cite{Cheng1989,Griffiths1999,Jackson1999}.
Since the magnet (according to the experiment) is falling with a constant velocity, 
it is immediately realized that the loss in mechanical potential 
energy of the magnet must be equal to the electrical resistive losses inside the metal cylinder. 

This experiment is oftenly conducted at high school level and in basic courses at the universities, etc., and is nowadays readily
accessible by everyone due to the availability of cheap super strong neodymium permanent magnets.
This type of rare-earth magnets are typically made of an alloy of neodymium, iron and boron ($\textrm{Nd}_{2}\textrm{Fe}_{14}\textrm{B}$),
and the sintered neodymium magnets are currently the most powerful permanent magnets that are 
commercially available \cite{Fraden2010}. The manufacturers grade these magnets in a scale ranging from N35 up to N52 
where a higher value indicates a stronger magnet. 
The super strong neodymium magnets make the experiment quite spectacular together with at thick metal cylinder made of copper or aluminum
(or any other non-magnetic and highly conductive material).
In the laboratory, an efficient way of significantly increasing the conductivity of the cylinder 
is to cool it down by using liquid nitrogen, or somewhat less spectacular by using dry ice or simply cooling it in a freezer.

Neodymium magnets are graded according to their maximum density of magnetic energy.
Other commonly used measures are the remanence of the magnet (the remaining magnetic flux density when there is no external field applied)
and the coercivity (the material's resistance to becoming demagnetized).
Even though these quantities are very useful to characterize the magnetic properties of the magnetic material, 
these parameters are non-trivially connected to the experimental set-up as described above, \ie to the physical dimensions of the magnet and of the cylinder. 
Furthermore, since the manufacturers specify these quantities based on test objects with different sizes their parameter values are most likely 
subjected to dimensional effects that depend on the geometry of the test object.
In the present context it is therefore the {\em magnetization} (in Amp\`{e}res per meter) that is the natural quantity to quantify the strength of the magnet
and which can be given a simple and rigorous interpretation in terms of the present experiment and where the dimensional effect is taken into full account.

This experiment has previously been analyzed in \cite{Saslow1992,Levin+etal2006a,Levin+etal2006b,Partovi+Morris2006}
where relevant approximations have been addressed.
It is easy to see that the quasi-static approximation (neglecting Maxwells displacement current and leading 
to a diffusion equation) is accurate for all practical purposes of this experiment, see \eg \cite{Partovi+Morris2006}.
In \cite{Saslow1992,Levin+etal2006a} the magnet is approximated by using a dipole source and the cylinder is assumed to be thin.
In \cite{Levin+etal2006b} the experiment is analyzed for the case with a superconducting pipe and 
a discussion is given regarding the limit of vanishing resistivity. In \cite{Saslow1992,Levin+etal2006a,Levin+etal2006b}
are also discussed the validity of neglecting the self-inductance of the cylinder (the induced currents). When applicable,
this approximation will greatly simplify the analysis and allow for closed form solutions.
Interestingly, even though this approximation usually is valid with strong magnets and normal conductors,
it turns out that the self-induction of the cylinder can not be ignored in the limit of vanishing resistivity. 
Ultimately, there will always be a skin-effect and a screening of the magnetic field which will cause
the magnet to fall freely \cite{Levin+etal2006b,Partovi+Morris2006}. This is also coherent with the outcome of the experiment if it would be
performed with a superconducting tube \cite{Levin+etal2006b}.
In \cite{Partovi+Morris2006} is finally given a complete and rigorous theory as a basis for a numerical solution of the diffusion equation, taking
the geometry of the magnet and of the cylinder as well as the self-induction (induced currents) into full account. 
In \cite{Partovi+Morris2006} is also given a quantitative analysis and arguments for the rapid vanishing of edge effects.
The aim of the present contribution is to provide a simple physical law and an explicit solution for the important case when the magnet is very strong
and the self-induction of the cylinder can be ignored.  The theory is general in that the geometry of the magnet as well as of the cylinder is included 
in the model.

The analysis presented here exploits cylindrical symmetries and relies on several well-known
properties regarding cylindrical Bessel functions and related Green's function expansions 
\cite{Watson1966,Collin1991,Jackson1999,Olver+etal2010,Arfken+etal2013,Bostrom+Kristensson+Strom1991}, and it provides finally an exact 
analytical expression for the motion of the falling magnet.
The analysis reveals the simple physical law for stationary motion 
\begin{equation}\nonumber
\sigma M^2v=-\frac{mg}{\mu_0^2C},
\end{equation}
where $\sigma$ is the conductivity of the metal cylinder, $M$ 
the magnetization of the magnet, $v$ the velocity of the fall,
$m$ the mass of the magnet, $g$ the gravitational acceleration, $\mu_0$ 
the permeability of vacuum and where $C$ is a structure constant that depends on the geometrical dimensions 
of the magnet and of the cylinder. The transient motion is exponential with time constant $\tau=m/(\sigma M^2\mu_0^2C)$.
The analysis is based on the assumption that the self-induction of the cylinder can be neglected.
It can be shown that this approximation is valid when the final velocity $v$ is much smaller
than the characteristic recession velocity $v_0=2/(\mu_0\sigma d)$ where $d$ is the thickness of the cylinder, 
see \cite{Saslow1992,Levin+etal2006a,Levin+etal2006b}.
This is usually the case with strong magnets, reasonable thick cylinders and ordinary metals (such as copper and aluminum, etc.).
A more precise determination of the validity or accuracy of the simplified theory can be obtained by performing
a complete numerical solution for the particular problem at hand as described in \cite{Partovi+Morris2006}.

Even though the analysis here is based solely on cylindrical symmetries, it is expected that very similar physical laws may
be derived for more general magnet geometries. In particular, an interesting approach for future work would be to expand the magnetic field
of the magnet in a multipole expansion (dipole, quadrupole, etc) and to employ proper transformations 
between spherical and cylindrical expansions \cite{Bostrom+Kristensson+Strom1991} to derive a more general theory.
Another interesting future extension of the theory would be to asses the limits of validity for the quasi-static approximation
by developing a rigorous solution to the full Maxwell equations based on a complete wave guide (TE/TM) theory.

The application of the presented theory is illustrated by identifying the magnetization of a magnet when
the conductivity of the cylinder and the velocity of the fall is known. A validation of the theory is obtained by demonstrating
the consistency of this identification based on different measurement cylinders.
The application of the theory is also illustrated by estimating the conductivity (and hence the temperature) of the metal cylinders based on
different temperature scenarios.

\section{Analytical solution}

\subsection{Problem formulation}
\begin{figure}[htb]
\begin{picture}(50,110)
\put(100,0){\makebox(50,110){\includegraphics[width=4.5cm]{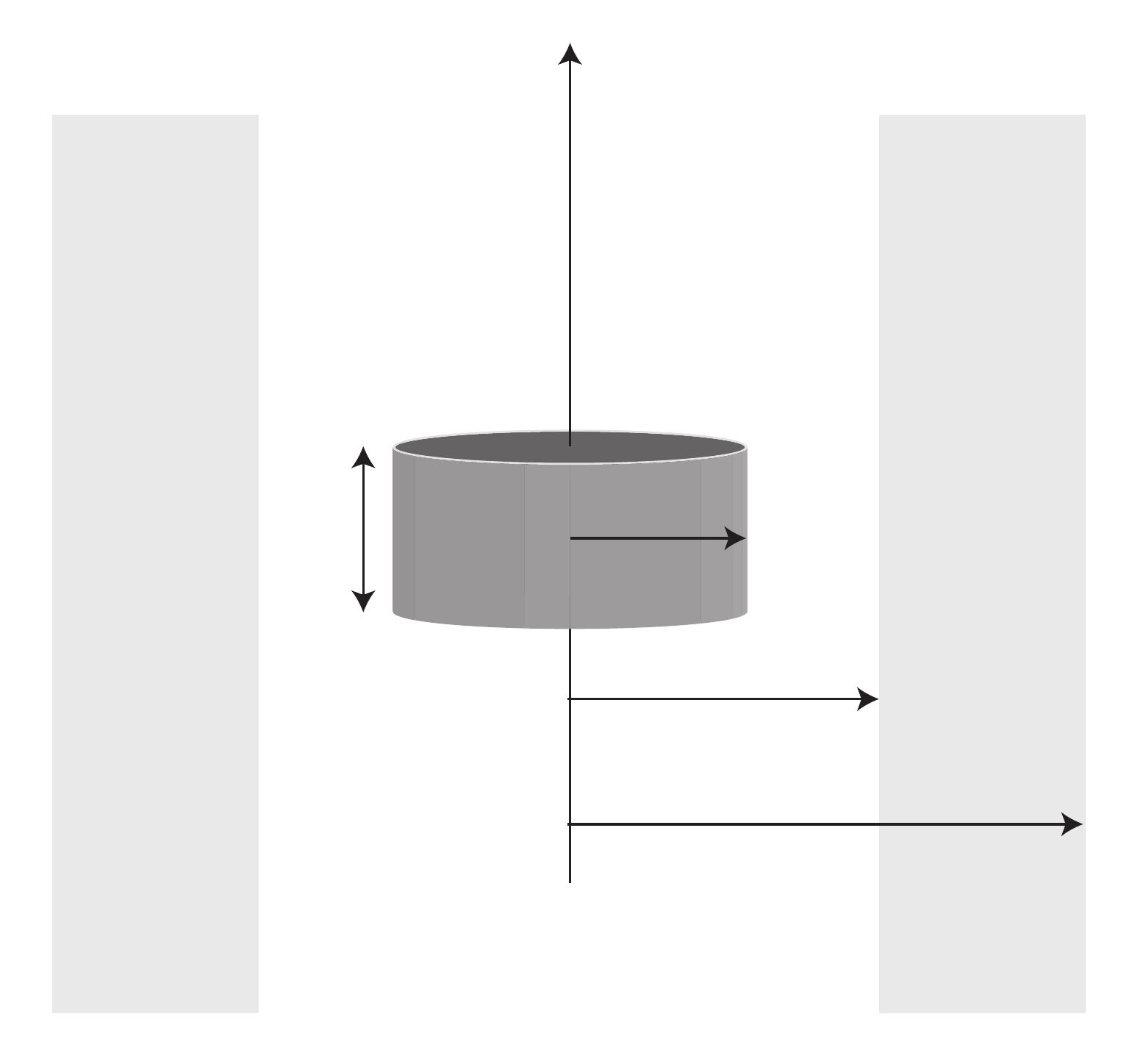}}} 
\put(133,106){\scriptsize $z$}
\put(133,58){\scriptsize  $a$}
\put(94,54){\scriptsize  $h$}
\put(149,41){\scriptsize  $\rho_1$}
\put(149,26){\scriptsize  $\rho_2$}
\end{picture}
\caption{Geometry of the problem.}
\label{fig:geometry}
\end{figure}
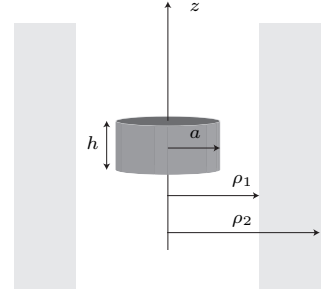

 A cylindrically shaped disc magnet with large homogeneous magnetization $M$\unit{[Am^{-1}]}, 
 mass $m$\unit{[kg]}, height $h$\unit{[m]} and radius $a$\unit{[m]} is placed inside a metal cylinder
 as depicted in Figure \ref{fig:geometry}. It is assumed that the cylinder is infinitely long and hence that edge effects can be neglected.
 The cylindrical coordinates are denoted $(\rho,\phi,z)$ and the corresponding unit vectors
 $(\hat{\vec{\rho}},\hat{\vec{\phi}},\hat{\vec{z}})$. The radius vector is $\vec{r}=\rho\hat{\vec{\rho}}+z\hat{\vec{z}}$.
 The cartesian coordinates are denoted $(x,y,z)$ and the corresponding unit vectors
 $(\hat{\vec{x}},\hat{\vec{y}},\hat{\vec{z}})$.
 
 The inner radius of the metal cylinder is $\rho_1$\unit{[m]} and the outer radius $\rho_2$\unit{[m]},
 and the conductivity 
 of the metal is $\sigma$\unit{[Sm^{-1}]}. 
 The electromagnetic response of the magnet is neglected 
 (the relative permittivity and permeability are assumed to be $\epsilon=\mu=1$), except that it exhibits
 a very strong homogeneous magnetization (dipole moment per unit volume) with a magnetization vector $\vec{M}=M\hat{\vec{z}}$.
 It is observed that this situation is equivalent to placing a cylindrically shaped sheat surface current in vacuum having a current density
 $\vec{J}_{\rm s}=-\hat{\vec{\rho}}\times\vec{M}=M\hat{\vec{\phi}}$ at the corresponding magnet surface
 at radius $a$, and that there are no other equivalent currents emanating from
 the top or bottom of the magnet, see \eg \cite{Cheng1989,Griffiths1999,Jackson1999}.
 
 The magnet is placed in a gravity field and experiences the force $-mg\hat{\vec{z}}$ where $g=9.81$\unit{ms^{-2}}.
 The position of the center of the magnet is denoted $z_0(t)$ where $t$ is the time. The magnet is released
 at $t=0$ with initial position $z_0(0)=0$ and velocity $\dot{z}_0(0)=0$, and where the dot denotes a differentiation with respect to the time. 
 The motion of the magnet will be determined for $t\geq 0$.
 
\subsection{The quasi-static approximation}\label{sect:quasi-static}
Let $\vec{E}(\vec{r},t)$, $\vec{D}(\vec{r},t)$, $\vec{H}(\vec{r},t)$, $\vec{B}(\vec{r},t)$ and $\vec{J}(\vec{r},t)$
denote the electric field intensity, the electric flux density, the magnetic field intensity, the magnetic flux density
and the electric current density, respectively, see \eg \cite{Cheng1989,Griffiths1999,Jackson1999}.
The Maxwell equations under the quasi-static approximation are given by 
\begin{equation}\label{eq:Maxwell1}
\left\{\begin{array}{l}
\nabla\times \vec{E}(\vec{r},t)=-\displaystyle\frac{\partial }{\partial t}\vec{B}(\vec{r},t), \vspace{0.2cm}\\
\nabla\times\vec{H}(\vec{r},t)=\vec{J}(\vec{r},t),
\end{array}\right.
\end{equation}
where the displacement current $\frac{\partial }{\partial t}\vec{D}(\vec{r},t)$ has been neglected
and there are no retarded potentials and no wave propagation phenomena \cite{Jackson1999}.
If the conduction current $\sigma\vec{E}(\vec{r},t)$ is included in the source term above a diffusion equation is obtained \cite{Jackson1999,Partovi+Morris2006}.

Consider a conductive material with real relative permittivity $\epsilon_{\rm r}$, conductivity $\sigma$ and displacement current 
\begin{equation}\label{eq:Displacement}
\displaystyle\frac{\partial }{\partial t}\vec{D}(\vec{r},t)=\epsilon_0\epsilon_{\rm r}\frac{\partial }{\partial t}\vec{E}(\vec{r},t)
+\sigma\vec{E}(\vec{r},t),
\end{equation}
where the conduction current is perceived as part of the displacement current.
Under normal circumstances with a strong magnet, normal metals and (hence) reasonable low velocities the first term
of the displacement current above can be neglected, see \eg \cite{Partovi+Morris2006}. To see this, consider the time scale
$\tau=d/v$ (frequency scale $\omega=v/d$) as the magnet passes through the cylinder. The ratio between the dielectric displacement current
and the conduction current is then $\omega\epsilon_0\epsilon_{\rm r}/\sigma=v\epsilon_{\rm r}/(c_0\eta_0\sigma d)$ where $c_0$ and $\eta_0$ are
the velocity and impedance of vacuum, respectively. In normal circumstances this ratio will be very small and can safely be ignored \cite{Partovi+Morris2006}.

The conduction term in \eqref{eq:Displacement}
(and hence the diffusion, and the self-induction) can be neglected only if the time scale for diffusion $\tau_{d}=d^2\mu_0\sigma$
\cite{Jackson1999} is much smaller than the time scale for the fall $\tau$, \ie $d^2\mu_0\sigma\ll d/v$, or $v\ll 1/(\mu_0\sigma d)$.	
This assertion is equivalent to require that the skin-depth $\delta=\sqrt{2/(\mu_0\sigma\omega)}$ \cite{Jackson1999}
is much larger then the characteristic dimension $d$,
\ie $\delta\gg d$, or $v\ll v_0=2/(\mu_0\sigma d)$ where $v_0$ is the characteristic recession velocity \cite{Saslow1992,Levin+etal2006a,Levin+etal2006b}.

To neglect the conduction current $\sigma\vec{E}(\vec{r},t)$ the magnet must also be very strong,
so that the effect of the equivalent current source $\vec{J}(\vec{r},t)$
will dominate over the effect of the displacement current $\frac{\partial }{\partial t}\vec{D}(\vec{r},t)$.
As the magnet is getting stronger ($M$ is getting larger) a smaller induced current inside the cylinder will be required to oppose the acceleration.
Hence, as $M$ is getting larger, the equivalent source term $\vec{J}(\vec{r},t)$ in \eqref{eq:Maxwell1} will increase at the same rate as
the induced current $\sigma\vec{E}(\vec{r},t)$ will decrease.  The magnet will fall with less power loss and a lower speed.

\subsection{Amp\`{e}re's current law}\label{sect:Ampere}
The first step is to determine the magnetic field of the permanent magnet when it is at rest at $z=0$. 
Here, the source consists of a cylindrically shaped sheat surface current with current density
$\vec{J}_{\rm s}=-\hat{\vec{\rho}}\times\vec{M}=M\hat{\vec{\phi}}$ at the surface of the magnet 
at radius $a$.
The magnetic vector potential is hence given by
\begin{equation}\label{eq:Adef1}
\vec{A}(\vec{r})=\mu_0\int_{0}^{2\pi}\int_{-h/2}^{h/2}\frac{1}{4\pi |\vec{r}-\vec{r}^\prime|}\vec{J}_{\rm s}(\vec{r}^\prime)\diff S^\prime,
\end{equation}
where $\mu_0=4\pi\cdot 10^{-7}$\unit{Hm^{-1}} is the permeability of vacuum,
$\vec{J}_{\rm s}(\vec{r}^\prime)=M\hat{\vec{\phi}}^\prime$ and $\diff S^\prime=a\diff\phi^\prime\diff z^\prime$, \cf \eg \cite{Cheng1989,Griffiths1999,Jackson1999}.

The potential $\vec{A}(\vec{r})$ is then evaluated in the $x$-$z$ plane where $\phi=0$, and adequate symmetries are exploited to yield
\begin{equation}\label{eq:Adef2}
\vec{A}(\vec{r})=\mu_0aM\hat{\vec{\phi}}\int_{0}^{2\pi}\int_{-h/2}^{h/2}
\frac{1}{4\pi |\vec{r}-\vec{r}^\prime|}\cos\phi^\prime\diff\phi^\prime\diff z^\prime,
\end{equation}
where $\vec{A}(\vec{r})$ is axial-symmetric ($A_\phi$ is independent of $\phi$),
$\hat{\vec{\phi}}^\prime=-\hat{\vec{x}}\sin\phi^\prime+\hat{\vec{y}}\cos\phi^\prime$, and in the $x$-$z$ plane
the Green's function $1/4\pi |\vec{r}-\vec{r}^\prime|$ is even in $\phi^\prime$ and $\hat{\vec{y}}=\hat{\vec{\phi}}$.

Next, the free space Green's function is expanded in cylindrical Bessel functions as 
\begin{multline}\label{eq:Greens1}
\frac{1}{4\pi |\vec{r}-\vec{r}^\prime|}
=\frac{\iu}{8\pi}\int_{-\infty}^{\infty}\sum_{m=-\infty}^{\infty}\\
{\rm J}_{m}(\iu|\alpha|\rho_<){\rm H}_{m}^{(1)}(\iu|\alpha|\rho_>)\eu^{\iu m(\phi-\phi^\prime)}\eu^{\iu\alpha(z-z^\prime)}\diff\alpha,
\end{multline}
see \eg \cite{Collin1991,Jackson1999,Arfken+etal2013} and Appendix \ref{app:ExpGreen}.
Here, ${\rm J}_{m}(\cdot)$ and ${\rm H}_{m}^{(1)}(\cdot)$ are the regular Bessel functions and the 
Hankel functions of the first kind, respectively, both of order $m$, see \eg \cite{Watson1966,Jackson1999,Olver+etal2010,Arfken+etal2013}. 
The arguments above are defined by $\rho_<=\min\{\rho,\rho^\prime\}$ and $\rho_>=\max\{\rho,\rho^\prime\}$, 
the integration variable $\alpha$ is a real valued Fourier variable corresponding to a Fourier transformation along the longitudinal coordinate $z$,
and $|\cdot|$ denotes the absolute value, see Appendix \ref{app:ExpGreen}.

By inserting \eqref{eq:Greens1} into \eqref{eq:Adef2} and integrating over the $\phi^\prime$ and $z^\prime$ coordinates,
the magnetic potential is given by
\begin{multline}\label{eq:Adef3}
\vec{A}(\vec{r})=\mu_0ahM\hat{\vec{\phi}}\frac{\iu}{4}  \int_{-\infty}^{\infty}\\
{\rm J}_{1}(\iu|\alpha|\rho_<){\rm H}_{1}^{(1)}(\iu|\alpha|\rho_>)
\frac{\sin(\alpha h/2)}{\alpha h/2}\eu^{\iu\alpha z}\diff\alpha,
\end{multline}
where $\rho^\prime=a$, and
where the following integrals have been used
\begin{equation}
\left\{\begin{array}{l}
\displaystyle \int_{0}^{2\pi}\eu^{-\iu m\phi^\prime}\cos\phi^\prime\diff\phi^\prime=\pi(\delta_{m,1}+\delta_{m,-1}), \vspace{0.2cm} \\
\displaystyle \int_{-h/2}^{h/2}\eu^{-\iu \alpha z^\prime}\diff z^\prime=h\frac{\sin (\alpha h/2)}{\alpha h/2},
\end{array}\right.
\end{equation}
where $\delta_{m,n}$ denotes the Kronecker delta \cite{Arfken+etal2013}, and where the relations ${\rm C}_{-m}(\zeta)=(-1)^m{\rm C}_{m}(\zeta)$ have been used which are valid for any cylinder function of order $m$ \cite{Olver+etal2010}.

The magnetic flux density is given by $\vec{B}=\nabla\times\vec{A}$, and in cylindrical coordinates
\begin{equation}
B_z(\rho,z)=\frac{1}{\rho}\frac{\partial }{\partial\rho}\rho A_{\phi}(\rho,z).
\end{equation}
The following explicit results are obtained
\begin{multline}\label{eq:Bdef1}
B_z^{(1)}(\rho,z)=\mu_0ahM\frac{\iu}{4}  \int_{-\infty}^{\infty}\\
\frac{1}{\rho}\frac{\partial}{\partial\rho}\rho
{\rm J}_{1}(\iu|\alpha|\rho){\rm H}_{1}^{(1)}(\iu|\alpha| a)\frac{\sin(\alpha h/2)}{\alpha h/2}\eu^{\iu\alpha z}\diff\alpha,
\end{multline}
where $\rho<\rho^\prime=a$, and
\begin{multline}\label{eq:Bdef2}
B_z^{(2)}(\rho,z)=\mu_0ahM\frac{\iu}{4}  \int_{-\infty}^{\infty}\\
\frac{1}{\rho}\frac{\partial}{\partial\rho}\rho
{\rm H}_{1}^{(1)}(\iu|\alpha| \rho){\rm J}_{1}(\iu|\alpha| a)\frac{\sin(\alpha h/2)}{\alpha h/2}\eu^{\iu\alpha z}\diff\alpha,
\end{multline}
where $\rho>\rho^\prime=a$. The total flux in the $z$-direction through a circular surface parallel to the $x$-$y$ plane
with radius $\rho>a$ and with its center positioned at height $z$ (and $x=y=0$), is given by
\begin{multline}\label{eq:Phidef}
\Phi(\rho,z)=\int_{0}^{a}\int_{0}^{2\pi}B_z^{(1)}(\rho^\prime,z)\rho^\prime\diff\rho^\prime\diff\phi^\prime \\
+\int_{a}^{\rho}\int_{0}^{2\pi}B_z^{(2)}(\rho^\prime,z)\rho^\prime\diff\rho^\prime\diff\phi^\prime
=\int_{-\infty}^{\infty}F(\alpha,\rho)\eu^{\iu\alpha z}\diff\alpha,
\end{multline}
where
\begin{equation}\label{eq:Fdef}
F(\alpha,\rho)=\mu_0ahM\frac{\pi}{2}\rho
{\rm H}_{1}^{(1)}(\iu|\alpha| \rho)\iu{\rm J}_{1}(\iu|\alpha| a)\frac{\sin(\alpha h/2)}{\alpha h/2}.
\end{equation}
The functions ${\rm H}_{1}^{(1)}(\iu|\alpha| \rho)$ and $\iu{\rm J}_{1}(\iu|\alpha| a)$ are real 
valued (see the next section), and hence the function $F(\alpha,\rho)$ is real valued and even in the variable $\alpha$.

\subsection{Computational issues and asymptotics of integrands}\label{sect:compissues}
For computational purposes it is useful to employ the regular and the singular modified Bessel functions
${\rm I}_m(\zeta)$ and ${\rm K}_m(\zeta)$, respectively, which are defined by
\begin{equation}\label{eq:ImKmdef}
\left\{\begin{array}{l}
{\rm I}_m(\zeta)=\iu^{-m}{\rm J}_m(\iu\zeta), \vspace{0.2cm} \\
{\rm K}_m(\zeta)=\iu^{m+1}\frac{\pi}{2}{\rm H}_m^{(1)}(\iu\zeta),
\end{array}\right.
\end{equation}
and which are real valued for real arguments $\zeta$, see \cite{Olver+etal2010,Arfken+etal2013}. 
For analysis purposes it is also useful to employ the following small argument asymptotics
\begin{equation}\label{eq:smallargas}
\left\{\begin{array}{l}
{\rm J}_m(\zeta)\sim\frac{1}{m!}\left(\frac{\zeta}{2}\right)^m,  \vspace{0.2cm} \\
{\rm H}_0^{(1)}(\zeta)\sim\iu\frac{2}{\pi}\ln\frac{\zeta}{2},\vspace{0.2cm} \\
{\rm H}_m^{(1)}(\zeta)\sim\frac{-\iu(m-1)!}{\pi}\left(\frac{2}{\zeta}\right)^m,
\end{array}\right.
\left\{\begin{array}{l}
{\rm I}_m(\zeta)\sim\frac{1}{m!}\left(\frac{\zeta}{2}\right)^m,  \vspace{0.2cm} \\
{\rm K}_0(\zeta)\sim -\ln\frac{\zeta}{2},\vspace{0.2cm} \\
{\rm K}_m(\zeta)\sim \frac{(m-1)!}{2}\left(\frac{2}{\zeta}\right)^m,
\end{array}\right.
\end{equation}
as $\zeta\rightarrow 0$, where $m\geq 0$ in the first line above and $m\geq 1$ in the third, \cf \cite{Olver+etal2010,Arfken+etal2013}.

To compute the magnetic flux density $B_z(\rho,z)$ for $\rho=0$ and $z=0$ based on \eqref{eq:Bdef1}, it is 
noted that the term $\frac{1}{\rho}\frac{\partial}{\partial\rho}\rho{\rm J}_{1}(\iu|\alpha|\rho)\sim\iu|\alpha|$
as $\rho\rightarrow 0$, and hence that
\begin{equation}\label{eq:Bdef1rho0z0}
B_z(0,0)=\mu_0ahM\frac{1}{\pi} \int_{0}^{\infty}
\alpha {\rm K}_{1}(\alpha a)\frac{\sin(\alpha h/2)}{\alpha h/2}\diff\alpha,
\end{equation}
where \eqref{eq:ImKmdef} has been used, as well as the fact that the integrand is even.  By using \eqref{eq:smallargas}, it is seen that
the integrand in \eqref{eq:Bdef1rho0z0} approaches the value $1/a$ as $\alpha\rightarrow 0$.

The large argument asymptotics of the regular Bessel functions and of the Hankel functions of the first kind
are given by
\begin{equation}\label{eq:largeargas}
\left\{\begin{array}{l}
{\rm J}_m(\zeta)\sim\sqrt{\frac{2}{\pi \zeta}}\cos(\zeta-\frac{1}{2}m\pi-\frac{1}{4}\pi), \vspace{0.2cm} \\
{\rm H}_m^{(1)}(\zeta)\sim\sqrt{\frac{2}{\pi \zeta}}\eu^{\iu(\zeta-\frac{1}{2}m\pi-\frac{1}{4}\pi)}
\left(1+\iu\frac{4m^2-1}{8\zeta} \right), 
\end{array}\right.
\end{equation}
as $\zeta\rightarrow\infty$, see  \cite{Olver+etal2010,Arfken+etal2013}. Hence, for large $\alpha$ it is concluded that the integrand 
in \eqref{eq:Bdef1rho0z0} decays with an exponential factor $\eu^{-\alpha a}$ as $\alpha\rightarrow\infty$.

\subsection{Faraday's law of induction}
The integral form of Faraday's law of induction \cite{Cheng1989,Griffiths1999,Jackson1999} is given by
\begin{equation}
\int_{\cal C}\vec{E}(\vec{r},t)\cdot\diff\vec{r}=\int_{\cal S}-\frac{\partial}{\partial t}\vec{B}(\vec{r},t)\cdot\diff\vec{S},
\end{equation}
where ${\cal S}$ is the circular surface with radius $\rho$ positioned at height $z$ and ${\cal C}$ its (right-handed) contour. 
The quasi-static approximation is now incorporated by letting $\Phi(\rho,z-z_0(t))$ describe the magnetic flux
of the moving magnet at position $z_0(t)$ and where $\Phi(\rho,\cdot)$ is given by \eqref{eq:Phidef}.
By further exploiting the axial symmetries using $\vec{E}(\vec{r},t)=\hat{\vec{\phi}}E_{\phi}(\rho,z,t)$, 
the Faraday's law of induction yields
\begin{multline}
E_{\phi}(\rho,z,t)2\pi\rho=-\frac{\partial}{\partial t}\Phi(\rho,z-z_0(t)) \\
=-\frac{\partial}{\partial t}\int_{-\infty}^{\infty}F(\alpha,\rho)\eu^{\iu\alpha (z-z_0(t))}\diff\alpha,
\end{multline}
or
\begin{equation}\label{eq:Ephidef1}
E_{\phi}(\rho,z,t)=\frac{\dot{z}_0(t)}{2\pi\rho}
\int_{-\infty}^{\infty}\iu\alpha F(\alpha,\rho)\eu^{\iu\alpha (z-z_0(t))}\diff\alpha.
\end{equation}
The induced current density is obtained as
\begin{equation}\label{eq:Jphidef1}
J_{\phi}(\rho,z,t)=\sigma E_{\phi}(\rho,z,t),
\end{equation}
and it is noted that $J_{\phi}(\rho,z,t)$ is a real valued function that is odd in the variable $z-z_0(t)$.
Hence, the induced current is circulating in opposite directions above and below the center of the magnet,
all in accordance to Lenz's law \cite{Cheng1989,Griffiths1999,Jackson1999}.

The total induced current for $z>z_0(t)$ is given by
\begin{multline}\label{eq:Idisp1}
I_{\rm ind}(t)=\int_{\rho_1}^{\rho_2}\int_{z_0(t)}^{\infty}J_{\phi}(\rho,z,t)\diff\rho\diff z \\
=\sigma\frac{\dot{z}_0(t)}{2\pi}\int_{-\infty}^{\infty}\iu\alpha \int_{\rho_1}^{\rho_2}\frac{1}{\rho}F(\alpha,\rho)\diff\rho
\int_{0}^{\infty}\eu^{\iu\alpha z}\diff z\diff\alpha \\
=\sigma\frac{\dot{z}_0(t)}{4}\mu_0ahM\int_{-\infty}^{\infty}
\frac{1}{|\alpha|}\frac{\sin(\alpha h/2)}{\alpha h/2}\\
{\rm J}_{1}(\iu|\alpha| a) 
\left( {\rm H}_{0}^{(1)}(\iu|\alpha| \rho_2)-{\rm H}_{0}^{(1)}(\iu|\alpha| \rho_1)\right)\diff\alpha.
\end{multline}
The relation \eqref{eq:Fdef} was used in the derivation above, as well as the identity
${\rm H}_{1}^{(1)}(\zeta)=-\frac{\partial}{\partial\zeta}{\rm H}_{0}^{(1)}(\zeta)$,
and the Fourier integral
\begin{equation}
\int_{0}^{\infty}\eu^{\iu\alpha z}\diff z=\frac{1}{-\iu\alpha}+\pi\delta(\alpha),
\end{equation}
which should be interpreted in the sense of distributions \cite{Zemanian1965}.

By using the modified Bessel functions defined in \eqref{eq:ImKmdef}, the total induced current \eqref{eq:Idisp1}
can also be written
\begin{multline}\label{eq:Idisp2}
I_{\rm ind}(t)= 
\sigma\dot{z}_0(t)\mu_0ahM\frac{1}{\pi}\int_{0}^{\infty}\frac{1}{\alpha}\frac{\sin(\alpha h/2)}{\alpha h/2}
\\ {\rm I}_{1}(\alpha a) 
\left( {\rm K}_{0}(\alpha \rho_2)-{\rm K}_{0}(\alpha \rho_1)\right)\diff\alpha.
\end{multline}

\subsection{Power loss and the velocity of the fall}\label{sect:powerloss}
The resistive power loss in the metal can be computed based on
Poyntings theorem \cite{Jackson1999} as
\begin{equation}
P_{\rm loss}(t)=\sigma\int_{\rho_1}^{\rho_2}\int_{0}^{2\pi}\int_{-\infty}^{\infty}\left| E_{\phi}(\rho,z,t)\right|^2\rho\diff\rho\diff\phi\diff z,
\end{equation}
where $E_{\phi}(\rho,z,t)$ is given by \eqref{eq:Ephidef1}. It follows that
\begin{multline}
P_{\rm loss}(t)=\sigma 2\pi \\
\int_{\rho_1}^{\rho_2}\int_{-\infty}^{\infty}\left(\frac{\dot{z}_0(t)}{2\pi\rho} \right)^2
\int_{-\infty}^{\infty}\iu\alpha F(\alpha,\rho)\eu^{\iu\alpha (z-z_0(t))}\diff\alpha \\
\int_{-\infty}^{\infty}-\iu\alpha^\prime F(\alpha^\prime,\rho)\eu^{-\iu\alpha^\prime (z-z_0(t))}\diff\alpha^\prime\rho\diff\rho\diff z\\
=\sigma 2\pi 
\int_{\rho_1}^{\rho_2}\left(\frac{\dot{z}_0(t)}{2\pi\rho} \right)^2
\int_{-\infty}^{\infty}\int_{-\infty}^{\infty}\alpha F(\alpha,\rho) 
\alpha^\prime F(\alpha^\prime,\rho)\\
\int_{-\infty}^{\infty}\eu^{\iu(\alpha-\alpha^\prime)(z-z_0(t))}\diff z
\diff\alpha\diff\alpha^\prime\rho\diff\rho\\
=\sigma \dot{z}^2_0(t)\int_{-\infty}^{\infty}\int_{\rho_1}^{\rho_2}\frac{1}{\rho^2}\alpha^2F^2(\alpha,\rho)\rho\diff\rho\diff\alpha,
\end{multline}
where the distributional relation $\int_{-\infty}^{\infty}\eu^{\iu(\alpha-\alpha^\prime)(z-z_0(t))}\diff z=2\pi\delta(\alpha-\alpha^\prime)$
has been used. By inserting \eqref{eq:Fdef} into the last line above, the following result is obtained
\begin{equation}\label{eq:Plosst}
P_{\rm loss}(t)=\sigma M^2 \dot{z}^2_0(t)\mu_0^2C,
\end{equation}
where $C$ (in units\unit{[m^3]}) is a structure constant that depends on the geometrical parameters $(a,h,\rho_1,\rho_2)$ 
of the magnet and of the cylinder, and
which is explicitly given by
\begin{multline}\label{eq:Cdef1}
C=a^2h^2\frac{\pi^2}{4}\int_{-\infty}^{\infty}\alpha^2\frac{\sin^2(\alpha h/2)}{(\alpha h/2)^2}
|{\rm J}_{1}(\iu|\alpha| a)|^2 \\
\int_{\rho_1}^{\rho_2}({\rm H}_{1}^{(1)}(\iu|\alpha|\rho))^2\rho\diff\rho\diff\alpha.
\end{multline}
The inner integral can be computed explicitly by use of the so called Lommel integral \cite{Watson1966,Olver+etal2010,Arfken+etal2013} 
yielding
\begin{multline}
\int_{\rho_1}^{\rho_2}({\rm H}_{1}^{(1)}(\iu|\alpha|\rho))^2\rho\diff\rho \\
=\frac{1}{2}\rho_2^2\left[({\rm H}_1^{(1)}(\iu|\alpha|\rho_2))^2-{\rm H}_0^{(1)}(\iu|\alpha|\rho_2){\rm H}_2^{(1)}(\iu|\alpha|\rho_2) \right] \\
-\frac{1}{2}\rho_1^2\left[({\rm H}_1^{(1)}(\iu|\alpha|\rho_1))^2-{\rm H}_0^{(1)}(\iu|\alpha|\rho_1){\rm H}_2^{(1)}(\iu|\alpha|\rho_1) \right].
\end{multline}

For computational purposes it is adequate to employ the modified Bessel functions defined in \eqref{eq:ImKmdef},
yielding
\begin{multline}\label{eq:Cdef2}
C=a^2h^2\int_{0}^{\infty}\alpha^2\frac{\sin^2(\alpha h/2)}{(\alpha h/2)^2}
{\rm I}_{1}^{2}(\alpha a) \\
\frac{\pi^2}{2}\int_{\rho_1}^{\rho_2}({\rm H}_{1}^{(1)}(\iu\alpha\rho))^2\rho\diff\rho\diff\alpha,
\end{multline}
where
\begin{multline}
\frac{\pi^2}{2}\int_{\rho_1}^{\rho_2}({\rm H}_{1}^{(1)}(\iu\alpha\rho))^2\rho\diff\rho \\
=\rho_2^2\left[({\rm K}_1(\alpha\rho_2))^2-{\rm K}_0(\alpha\rho_2){\rm K}_2(\alpha\rho_2) \right] \\
-\rho_1^2\left[({\rm K}_1(\alpha\rho_1))^2-{\rm K}_0(\alpha\rho_1){\rm K}_2(\alpha\rho_1) \right].
\end{multline}
By using \eqref{eq:smallargas}, it is seen that
the integrand in \eqref{eq:Cdef2}  approaches the value $0$ as $\alpha\rightarrow 0$. 
For large $\alpha$,  \eqref{eq:ImKmdef} and \eqref{eq:largeargas} are used to conclude that the integrand 
in \eqref{eq:Cdef2} decays with a dominating exponential factor $\eu^{-2\alpha (\rho_1-a)}$ as $\alpha\rightarrow\infty$.

Experiments show that the magnet very quickly assumes a constant velocity as it falls through the metal cylinder.
When there is a constant velocity $v$ (in units\unit{[ms^{-1}]}) there is no power exchange associated with an acceleration of the magnet,
and all the resistive losses must be attributed to the loss in mechanical potential energy.
Based on \eqref{eq:Plosst} this gives the power balance equation
\begin{equation}
\sigma M^2 v^2\mu_0^2C=-mgv,
\end{equation}
yielding the simple physical law for stationary motion
\begin{equation}\label{eq:powerrel}
\sigma M^2 v=-\frac{mg}{\mu_0^2 C},
\end{equation}
where the structure constant $C$ has been defined in \eqref{eq:Cdef1} and \eqref{eq:Cdef2}.

The resulting velocity $v$ given by \eqref{eq:powerrel} can now be inserted into \eqref{eq:Idisp2} to yield the total induced current
\begin{multline}\label{eq:Idisp3}
I_{\rm ind}= -\frac{mgah}{\mu_0 M C}\frac{1}{\pi}\int_{0}^{\infty}\frac{1}{\alpha}\frac{\sin(\alpha h/2)}{\alpha h/2} \\
{\rm I}_{1}(\alpha a) \left( {\rm K}_{0}(\alpha \rho_2)-{\rm K}_{0}(\alpha \rho_1)\right)\diff\alpha.
\end{multline}
It is noted that the induced current in \eqref{eq:Idisp3} is independent of the conductivity $\sigma$, as expected.
The integrand in \eqref{eq:Idisp3} approaches the value $-(a/2)\ln(\rho_2/\rho_1)$ as $\alpha\rightarrow 0$, and 
for large $\alpha$  the integrand decays with a dominating exponential factor $\eu^{-\alpha (\rho_1-a)}$ as $\alpha\rightarrow\infty$.

\subsection{Equation of motion}
To determine the motion of the falling magnet the induced magnetic reaction forces acting on the magnet must be derived.
Hence, the induced magnetic potential is calculated from the induced  current $J_{\phi}(\rho,z,t)$ as
\begin{multline}\label{eq:AAdef1}
\vec{A}^{\rm ind}(\vec{r},t)=\mu_0\int_{\rho_1}^{\rho_2}\int_{0}^{2\pi}\int_{-\infty}^{\infty} \\
\frac{1}{4\pi |\vec{r}-\vec{r}^\prime|}J_{\phi}(\rho^\prime,z^\prime,t)\hat{\vec{\phi}}^\prime\rho^\prime\diff\rho^\prime\diff\phi^\prime\diff z^\prime.
\end{multline}
The field $\vec{A}^{\rm ind}(\vec{r},t)$ is then evaluated in the $x$-$z$ plane where $\phi=0$, and adequate symmetries are exploited to yield
\begin{multline}\label{eq:AAdef2}
\vec{A}^{\rm ind}(\vec{r},t)=\mu_0\hat{\vec{\phi}}\int_{\rho_1}^{\rho_2}\int_{0}^{2\pi}\int_{-\infty}^{\infty} \\
\frac{1}{4\pi |\vec{r}-\vec{r}^\prime|}J_{\phi}(\rho^\prime,z^\prime,t)\cos\phi^\prime\rho^\prime\diff\rho^\prime\diff\phi^\prime\diff z^\prime,
\end{multline}
where $\vec{A}^{\rm ind}(\vec{r},t)$ is axial-symmetric, $\hat{\vec{\phi}}^\prime=-\hat{\vec{x}}\sin\phi^\prime+\hat{\vec{y}}\cos\phi^\prime$, 
and in the $x$-$z$ plane the Green's function $1/4\pi |\vec{r}-\vec{r}^\prime|$ is even in $\phi^\prime$ and $\hat{\vec{y}}=\hat{\vec{\phi}}$.

By inserting the free space Green's function expansion \eqref{eq:Greens1} into \eqref{eq:AAdef2} 
and integrating over the $\phi^\prime$ coordinate,
the magnetic potential is given by
\begin{multline}\label{eq:AAdef3}
\vec{A}^{\rm ind}(\vec{r},t)=\mu_0\frac{\iu}{4}\hat{\vec{\phi}} \int_{-\infty}^{\infty} \int_{\rho_1}^{\rho_2}\int_{-\infty}^{\infty}\\
{\rm J}_{1}(\iu|\alpha|\rho){\rm H}_{1}^{(1)}(\iu|\alpha|\rho^\prime)J_{\phi}(\rho^\prime,z^\prime,t)
\eu^{\iu\alpha (z-z^\prime)}\rho^\prime\diff\rho^\prime\diff z^\prime\diff\alpha,
\end{multline}
where $\rho<\rho_1$, \cf section \ref{sect:Ampere} for further details.
Next, by inserting \eqref{eq:Fdef}, \eqref{eq:Ephidef1} and \eqref{eq:Jphidef1} into the expression above, and integrating over
the $z^\prime$ coordinate, the magnetic potential becomes
\begin{multline}
A_{\phi}^{\rm ind}(\rho,z,t)=\mu_0^2\left(\frac{\iu\pi}{2}\right)^2\sigma\frac{\dot{z}_0(t)}{2\pi}ahM \\
\int_{-\infty}^{\infty}\iu\alpha {\rm J}_{1}(\iu|\alpha|\rho){\rm J}_{1}(\iu|\alpha| a)\frac{\sin(\alpha h/2)}{\alpha h/2}
\eu^{\iu\alpha (z-z_0(t))} \\
\int_{\rho_1}^{\rho_2}({\rm H}_{1}^{(1)}(\iu|\alpha|\rho^\prime))^2\rho^\prime\diff\rho^\prime\diff\alpha,
\end{multline}
where the relation $\int_{-\infty}^{\infty}\eu^{\iu(\alpha-\alpha^\prime)z^\prime}\diff z^\prime=2\pi\delta(\alpha-\alpha^\prime)$
has been used. The induced magnetic flux density at $\rho=a$ is given by
\begin{multline}
B_{\rho}^{\rm ind}(a,z,t)=-\frac{\partial}{\partial z}A_{\phi}^{\rm ind}(a,z,t) 
=\mu_0^2\frac{\pi^2}{4}\sigma\frac{\dot{z}_0(t)}{2\pi}ahM \\
\int_{-\infty}^{\infty}\alpha^2 \left|{\rm J}_{1}(\iu|\alpha|a)\right|^2\frac{\sin(\alpha h/2)}{\alpha h/2}
\eu^{\iu\alpha (z-z_0(t))} \\
\int_{\rho_1}^{\rho_2}({\rm H}_{1}^{(1)}(\iu|\alpha|\rho^\prime))^2\rho^\prime\diff\rho^\prime\diff\alpha.
\end{multline}
The magnetic force acting on the magnet is given by the Lorentz force equation
\begin{equation}
\vec{F}^{\rm mag}=\int_{0}^{2\pi}\int_{z_0(t)-h/2}^{z_0(t)+h/2}\vec{J}_{\rm s}(\vec{r})\times\vec{B}^{\rm ind}(\vec{r})\diff S,
\end{equation}
and hence
\begin{multline}
F_z^{\rm mag}=-M\int_{0}^{2\pi}\int_{z_0(t)-h/2}^{z_0(t)+h/2}B_{\rho}^{\rm ind}(a,z,t)a\diff\phi\diff z \\
=-2\pi Ma \int_{z_0(t)-h/2}^{z_0(t)+h/2}B_{\rho}^{\rm ind}(a,z,t)\diff z.
\end{multline}
By using
\begin{equation}
\int_{z_0(t)-h/2}^{z_0(t)+h/2}\eu^{\iu\alpha (z-z_0(t))}\diff z=h\frac{\sin(\alpha h/2)}{\alpha h/2},
\end{equation}
the final expression for the magnetic force is
\begin{equation}
F_z^{\rm mag}=-\sigma M^2\mu_0^2 C\dot{z}_0(t),
\end{equation}
where $C$ is the structure constant given by \eqref{eq:Cdef1} and \eqref{eq:Cdef2}.

The equation of motion is obtained by applying Newton's second law
\begin{equation}
-\sigma M^2 \mu_0^2C\dot{z}_0(t)-mg=m\ddot{z}_0(t),
\end{equation}
which has an exponential solution for $t\geq 0$
\begin{equation}
\left\{\begin{array}{l}
z_0(t)=g\tau\left[\tau(1-\eu^{-t/\tau})-t \right], \vspace{0.2cm} \\
\dot{z}_0(t)=g\tau\left[\eu^{-t/\tau}-1 \right], \vspace{0.2cm} \\
\ddot{z}_0(t)=-g\eu^{-t/\tau},
\end{array}\right.
\end{equation}
where the time constant $\tau$ is given by
\begin{equation}\label{eq:taudef1}
\tau=\frac{m}{\sigma M^2\mu_0^2C}.
\end{equation}
The stationary velocity is 
\begin{equation}
v=-g\tau=-\frac{mg}{\sigma M^2\mu_0^2C},
\end{equation}
which is in agreement with \eqref{eq:powerrel}.

\section{Experimental validation and estimation}
\subsection{Experimental set-up and measurements}
Two neodymium permanent disc magnets of grade N45 and N42 have been tested together with two
metal cylinders made of aluminum and copper. These measurement objects are referred to here as
N45, N42, Al and Cu, respectively, and their structural parameters are listed in Table \ref{tab:tab1}.

\begin{table}[htb]\centerline{
\begin{tabular}{||c||c|c|c||}
\cline{1-4} \scriptsize
\scriptsize Magnet  & \scriptsize radius $a$\unit{[mm]} & \scriptsize height $h$\unit{[mm]} & \scriptsize mass $m$\unit{[g]}  \\
%\scriptsize    & \scriptsize \unit{[mm]}  & \scriptsize\unit{[mm]} &  \scriptsize\unit{[g]}     \\
\cline{1-4} \scriptsize N45    & \scriptsize$15$  & \scriptsize$20$ & \scriptsize$107$        \\
\cline{1-4} \scriptsize N42   & \scriptsize$17.5$ & \scriptsize$20$ & \scriptsize$144$       \\
\cline{1-4} 
\scriptsize Cylinder  & \scriptsize radius $\rho_1$\unit{[mm]} & \scriptsize radius $\rho_2$\unit{[mm]} & \scriptsize length $L$\unit{[mm]}  \\
\cline{1-4} \scriptsize Al    & \scriptsize$20$  & \scriptsize$30$ & \scriptsize$102$        \\
\cline{1-4} \scriptsize Cu   & \scriptsize$16.1$ & \scriptsize$17.5$ & \scriptsize$136$       \\
\cline{1-4} 
\end{tabular}}
\hspace{5mm}
\caption{Magnet and cylinder data.}\label{tab:tab1}
\end{table}

The experiments have been conducted in three different test cases denoted by N45-Al, N42-Al and N45-Cu,
and with the measurement cylinder in four different temperature scenarios: 
1) cylinder heated with boiling water (+100\degree\unit{C}) during a few minutes; 
2) cylinder in room temperature (+23\degree\unit{C}) for calibration; 
3) cylinder cooled in a freezer (-20\degree\unit{C}) for at least 8 hours and
4) cylinder cooled with liquid nitrogen (-196\degree\unit{C}) during a few minutes. 
The experiments were recorded on video and the time required for the magnet to fall the distance $L-h$ through the cylinder was measured
by time-stepping and visually inspecting the video.
The resulting measured timing data in the four temperature scenarios are summarized in Table \ref{tab:tab2}.

\begin{table}[htb]\centerline{
\begin{tabular}{||c||c|c|c|c||}
\cline{1-5} \scriptsize
\scriptsize Test case  & \scriptsize +100\degree C & \scriptsize +23\degree C & \scriptsize -20\degree C & \scriptsize -196\degree C     \\
\scriptsize timing   & \scriptsize $t_1$\unit{[s]}  &   \scriptsize$t_2$\unit{[s]}    &    \scriptsize $t_3$\unit{[s]}   &  \scriptsize  $t_4$\unit{[s]}     \\
\cline{1-5} \scriptsize N45-Al      & \scriptsize$0.90$ & \scriptsize$1.10$ & \scriptsize$1.30$  & \scriptsize$3.85$    \\
\cline{1-5} \scriptsize N42-Al     & \scriptsize$1.43$ & \scriptsize$1.73$ & \scriptsize$2.05$ & \scriptsize$5.98$     \\
\cline{1-5} \scriptsize N45-Cu     & \scriptsize$1.78$ & \scriptsize$2.00$  & \scriptsize$2.29$ & \scriptsize$9.35$   \\
\cline{1-5} 
\end{tabular}}
\hspace{5mm}
\caption{Measured timing data.}\label{tab:tab2}
\end{table}

\subsection{Computation and parameter estimation}
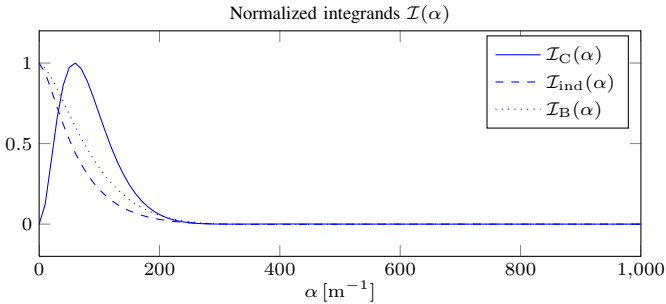
\begin{figure}[htb]
\begin{center}
%\begin{picture}(230,350)
\scriptsize
% This file was created by matlab2tikz v0.4.7 running on MATLAB 7.10.
% Copyright (c) 2008--2014, Nico Schlömer <nico.schloemer@gmail.com>
% All rights reserved.
% Minimal pgfplots version: 1.3
% 
% The latest updates can be retrieved from
%   http://www.mathworks.com/matlabcentral/fileexchange/22022-matlab2tikz
% where you can also make suggestions and rate matlab2tikz.
% 
\begin{tikzpicture}

\begin{axis}[%
width=8cm,
height=3cm,
scale only axis,
xmin=0,
xmax=1000,
xlabel style={yshift=0.3cm},
xlabel={$\alpha$\unit{[m^{-1}]}},
ymin=-0.2,
ymax=1.2,
title style={yshift=-0.2cm},
title={Normalized integrands ${\cal I}(\alpha)$},
legend style={draw=black,fill=white,legend cell align=left}
]
\addplot [color=blue,solid]
  table[row sep=crcr]{0	0\\
10	0.122951768061974\\
20	0.389307544525004\\
30	0.660857049942308\\
40	0.862524714212143\\
50	0.972793824819064\\
60	1\\
70	0.964346955514174\\
80	0.887616407546848\\
90	0.788454013335159\\
100	0.680908626204469\\
110	0.57458222214035\\
120	0.475418571524062\\
130	0.386625920252077\\
140	0.309508181090221\\
150	0.244128811682179\\
160	0.189803220810037\\
170	0.145444460313615\\
180	0.109794473491448\\
190	0.0815711693029083\\
200	0.0595562126905698\\
210	0.0426425614519526\\
220	0.0298556249935471\\
230	0.0203578150276716\\
240	0.0134431743974187\\
250	0.00852654210477151\\
260	0.00513014809755665\\
270	0.00286945750533134\\
280	0.00143936134921787\\
290	0.00060133354244952\\
300	0.000171865164428325\\
310	1.22918663957921e-05\\
320	2.00108682913424e-05\\
330	0.000121014359958364\\
340	0.000263628693875676\\
350	0.000413331848803212\\
360	0.000548517520337331\\
370	0.000657077791807717\\
380	0.000733684372730954\\
390	0.000777658752996347\\
400	0.000791332970678522\\
410	0.00077881419369987\\
420	0.000745077459971303\\
430	0.00069532140684927\\
440	0.000634531483904274\\
450	0.00056720390534002\\
460	0.000497191438156447\\
470	0.000427639053471452\\
480	0.000360983528409732\\
490	0.000298996326167063\\
500	0.00024285356253459\\
510	0.000193220653421919\\
520	0.000150342396915866\\
530	0.000114131842368357\\
540	8.4253403678903e-05\\
550	6.01973474840508e-05\\
560	4.13440890953217e-05\\
570	2.70177153664177e-05\\
580	1.65288753369399e-05\\
590	9.20768283686574e-06\\
600	4.42760169638754e-06\\
610	1.62147036261964e-06\\
620	2.90900388907884e-07\\
630	1.02796882894002e-08\\
640	4.2654960287111e-07\\
650	1.25582375370305e-06\\
660	2.27779172660605e-06\\
670	3.32871413307625e-06\\
680	4.29367685448874e-06\\
690	5.09863828711814e-06\\
700	5.70267904210775e-06\\
710	6.09075198694902e-06\\
720	6.2671335154444e-06\\
730	6.24969516918036e-06\\
740	6.06504802762349e-06\\
750	5.7445598389778e-06\\
760	5.32120545600614e-06\\
770	4.82718328029119e-06\\
780	4.2922124800212e-06\\
790	3.74241607401298e-06\\
800	3.19969195944803e-06\\
810	2.68147609863115e-06\\
820	2.20080801052162e-06\\
830	1.766617242241e-06\\
840	1.38415960810643e-06\\
850	1.05554284085925e-06\\
860	7.80292234507782e-07\\
870	5.55917361815907e-07\\
880	3.78450655349901e-07\\
890	2.42937306354188e-07\\
900	1.43863422375773e-07\\
910	7.55156398349856e-08\\
920	3.22704263459915e-08\\
930	8.81519484691036e-09\\
940	3.06188531496689e-10\\
950	2.47000598105836e-09\\
960	1.16567555010562e-08\\
970	2.48532953181529e-08\\
980	3.96649670935831e-08\\
990	5.42737904301642e-08\\
1000	6.73803696249008e-08\\
};
\addlegendentry{${\cal I}_{\rm C}(\alpha)$};

\addplot [color=blue,dashed]
  table[row sep=crcr]{0	1\\
10	0.938870004037618\\
20	0.834954882646272\\
30	0.724505456394045\\
40	0.619748291308762\\
50	0.525307531163433\\
60	0.442492203121134\\
70	0.37107572194953\\
80	0.31014509902823\\
90	0.258528535028892\\
100	0.215012258665107\\
110	0.178446666341064\\
120	0.147793133977759\\
130	0.122139128894589\\
140	0.100696859067604\\
150	0.0827939410620935\\
160	0.0678607819498272\\
170	0.0554172168974651\\
180	0.0450597134683888\\
190	0.0364497544550736\\
200	0.0293036208103273\\
210	0.0233835860507515\\
220	0.0184904276743979\\
230	0.0144571149264872\\
240	0.0111435191876815\\
250	0.00843199775865291\\
260	0.00622371480055759\\
270	0.00443557949645921\\
280	0.00299769828601838\\
290	0.00185125383286957\\
300	0.000946737547586308\\
310	0.000242474796104074\\
320	-0.0002966075964315\\
330	-0.000700013133651014\\
340	-0.000992540990792979\\
350	-0.00119501312086596\\
360	-0.00132488634954743\\
370	-0.00139676859438838\\
380	-0.00142285499246181\\
390	-0.00141329697600521\\
400	-0.00137651508776994\\
410	-0.00131946448555944\\
420	-0.00124786057392209\\
430	-0.00116637095895108\\
440	-0.00107877889977639\\
450	-0.000988122587099807\\
460	-0.000896813882141248\\
470	-0.000806739571959156\\
480	-0.000719347717636169\\
490	-0.000635721272710836\\
500	-0.000556640816202804\\
510	-0.000482637965996687\\
520	-0.000414040804739181\\
530	-0.000351012454022157\\
540	-0.000293583767134222\\
550	-0.000241680970866869\\
560	-0.000195148968486475\\
570	-0.00015377091551077\\
580	-0.000117284594451868\\
590	-8.53960418033604e-05\\
600	-5.77908182618916e-05\\
610	-3.41432598267315e-05\\
620	-1.41240016333967e-05\\
630	2.59397300329276e-06\\
640	1.63305397994032e-05\\
650	2.73947077319687e-05\\
660	3.60817964045935e-05\\
670	4.26713746300359e-05\\
680	4.74258373763226e-05\\
690	5.05895147951088e-05\\
700	5.23882214758064e-05\\
710	5.3029166639801e-05\\
720	5.27011569555199e-05\\
730	5.15750332300755e-05\\
740	4.98042905986671e-05\\
750	4.7525839144774e-05\\
760	4.48608682763183e-05\\
770	4.19157837706846e-05\\
780	3.87831912842695e-05\\
790	3.55429043861973e-05\\
800	3.22629588957072e-05\\
810	2.90006185434882e-05\\
820	2.5803359795637e-05\\
830	2.27098261245418e-05\\
840	1.97507441273765e-05\\
850	1.69497957182851e-05\\
860	1.43244421885102e-05\\
870	1.18866972694727e-05\\
880	9.6438474736018e-06\\
890	7.59911894998359e-06\\
900	5.75229089754662e-06\\
910	4.10025624612452e-06\\
920	2.63753086195756e-06\\
930	1.3567129738606e-06\\
940	2.48894862684049e-07\\
950	-6.95970878269686e-07\\
960	-1.48874800988557e-06\\
970	-2.14083360576932e-06\\
980	-2.66391258694302e-06\\
990	-3.06974708134641e-06\\
1000	-3.36999785925879e-06\\
};
\addlegendentry{${\cal I}_{\rm ind}(\alpha)$};

\addplot [color=blue,dotted]
  table[row sep=crcr]{0	1\\
10	0.970006251223512\\
20	0.910697838303007\\
30	0.838474292237519\\
40	0.76102173381674\\
50	0.682879690232032\\
60	0.606877938286032\\
70	0.534779520686116\\
80	0.467636248178987\\
90	0.406009967215614\\
100	0.350120678436175\\
110	0.299950351467372\\
120	0.255317800523583\\
130	0.215933593045673\\
140	0.181440627154122\\
150	0.151444131561755\\
160	0.125533708107355\\
170	0.103299314866679\\
180	0.0843426045517693\\
190	0.0682846955941072\\
200	0.0547712092430545\\
210	0.0434752240737124\\
220	0.0340986603688848\\
230	0.0263724987364505\\
240	0.0200561520070946\\
250	0.0149362414401622\\
260	0.0108249736768925\\
270	0.00755827090511073\\
280	0.00499377123053327\\
290	0.00300878766944223\\
300	0.00149829121590312\\
310	0.000372965078940523\\
320	-0.0004426373921956\\
330	-0.00101181004875399\\
340	-0.00138727442897365\\
350	-0.00161261811370583\\
360	-0.00172359082333632\\
370	-0.00174926190562366\\
380	-0.00171304554007097\\
390	-0.00163360178385135\\
400	-0.00152562269381364\\
410	-0.00140051333308441\\
420	-0.00126697763621424\\
430	-0.0011315189674296\\
440	-0.000998864846457929\\
450	-0.000872324803173833\\
460	-0.00075408970968304\\
470	-0.000645480268562803\\
480	-0.000547151641358988\\
490	-0.000459260506739312\\
500	-0.000381600161093875\\
510	-0.000313708628794986\\
520	-0.00025495414351695\\
530	-0.00020460180138537\\
540	-0.000161864674091093\\
550	-0.000125942206292588\\
560	-9.60483059766235e-05\\
570	-7.14311672006123e-05\\
580	-5.13865392950248e-05\\
590	-3.52658721618562e-05\\
600	-2.24805204803528e-05\\
610	-1.25029770135317e-05\\
620	-4.86592338439242e-06\\
630	8.40267668291642e-07\\
640	4.9710746142988e-06\\
650	7.83201885759047e-06\\
660	9.68320089696949e-06\\
670	1.07439501417231e-05\\
680	1.11974180439136e-05\\
690	1.11949929500762e-05\\
700	1.08604539301665e-05\\
710	1.02938114082274e-05\\
720	9.57480609274151e-06\\
730	8.76605569433285e-06\\
740	7.91585226464168e-06\\
750	7.06062258090282e-06\\
760	6.2270705884994e-06\\
770	5.43402513113688e-06\\
780	4.69401857286216e-06\\
790	4.0146228841568e-06\\
800	3.39956968317223e-06\\
810	2.84967988276971e-06\\
820	2.36362722695365e-06\\
830	1.93855829077108e-06\\
840	1.57058960958425e-06\\
850	1.25520060706789e-06\\
860	9.87538989046246e-07\\
870	7.6265332265318e-07\\
880	5.75665669513229e-07\\
890	4.21895415599117e-07\\
900	2.96943855773105e-07\\
910	1.96747655757466e-07\\
920	1.17608029858568e-07\\
930	5.62013358466865e-08\\
940	9.57579223072239e-09\\
950	-2.48618412528704e-08\\
960	-4.93665222076645e-08\\
970	-6.58796583601508e-08\\
980	-7.60563837288074e-08\\
990	-8.12938936269965e-08\\
1000	-8.27597231711598e-08\\
};
\addlegendentry{${\cal I}_{\rm B}(\alpha)$};

\end{axis}
\end{tikzpicture}%
%\end{picture}
\end{center}
\vspace{-5mm}
\caption{Numerical integration for the test case N45-Al at calibration temperature $T=+23\degree C$. 
The plots show the normalized integrands ${\cal I}_{\rm C}(\alpha)$, ${\cal I}_{\rm ind}(\alpha)$ and ${\cal I}_{\rm B}(\alpha)$  
in the numerical integration of $C$ defined by \eqref{eq:Cdef2}, $I_{\rm ind}$ defined by \eqref{eq:Idisp3}, and
$B_z(0,0)$ defined by \eqref{eq:Bdef1rho0z0}, respectively.
}
\label{fig:matfig21}
\end{figure}

The structure constant $C$ is first calculated based on \eqref{eq:Cdef2}, 
and the magnetization $M$ of the magnet is then identified by
using \eqref{eq:powerrel} together with data ($\sigma$ and $v$) corresponding to the calibration measurement at room temperature. 
The conductivity data at different temperatures for aluminum and copper have been
obtained from \cite{Lide2008}. Here, $\sigma=3.77\cdot10^7$\unit{Sm^{-1}} for aluminum at $+20\degree\unit{C}$ and
$\sigma=5.96\cdot10^7$\unit{Sm^{-1}} for copper at $+20\degree\unit{C}$, and which are adjusted to  $+23\degree\unit{C}$
by using the appropriate temperature coefficients \cite{Serway1998}.
Once the magnetization $M$ of the magnet has been identified, the induced current $I_{\rm ind}$
and the maximum magnetic flux density $B=B_z(0,0)$ can be calculated from 
\eqref{eq:Idisp3} and \eqref{eq:Bdef1rho0z0}, respectively.
The normalized integrands for computing $C$, $I_{\rm ind}$ and $B$ are shown in Figure \ref{fig:matfig21} for the test case N45-Al. 
It has been shown in section \ref{sect:compissues} and \ref{sect:powerloss} that these integrands are well behaved continuous functions 
that decay exponentially for large $\alpha$. 
Hence, a simple numerical integration scheme is used here
based on the composite Simpson's rule \cite{Dahlquist+Bjork1974}, and all the integrals evaluated in these examples 
converged to 4-5 digits based on $N=1001$ sample points and an integration interval $[0,1000]$.
The computer software MATLAB was used and the computation time for each test case was less than 1 second on a standard PC.
The resulting set of estimated parameters for the three test cases are summarized in Table~\ref{tab:tab3}
including the time constant $\tau$ given by \eqref{eq:taudef1} as well as the measured velocity $v$
used for calibration (and where $v=g\tau$).

\begin{table}[htb]\centerline{
\begin{tabular}{||c||c|c|c|c|c|c||}
\cline{1-7} \scriptsize
\scriptsize Test case  & \scriptsize $C$ & \scriptsize $M$  & \scriptsize $I_{\rm ind}$ & \scriptsize $B$  & \scriptsize $\tau$\unit{[ms]} 
&  \scriptsize $v$\unit{[cms^{-1}]}    \\
\scriptsize    & \scriptsize \unit{[mm^3]} &   \scriptsize \unit{[kAm^{-1}]}  &    \scriptsize \unit{[A]}  &  \scriptsize \unit{[T]}  &  \scriptsize at +23\degree\unit{C}
&  \scriptsize at +23\degree\unit{C}    \\
\cline{1-7} \scriptsize N45-Al      & \scriptsize$296$ & \scriptsize$899$ & \scriptsize$61$  & \scriptsize$0.63$ & \scriptsize$7.6$ & \scriptsize$7.5$   \\
\cline{1-7} \scriptsize N42-Al     & \scriptsize$647$ & \scriptsize$884$ & \scriptsize$54$ & \scriptsize$0.55$ & \scriptsize$4.8$ & \scriptsize$4.7$   \\
\cline{1-7} \scriptsize N45-Cu     & \scriptsize$193$ & \scriptsize$1003$  & \scriptsize$24$ & \scriptsize$0.70$ & \scriptsize$5.9$ & \scriptsize$5.8$ \\
\cline{1-7} 
\end{tabular}}
\hspace{5mm}
\caption{Estimated parameters of the neodymium magnets. The rightmost column shows the measured velocities at +23\degree\unit{C}.}
\label{tab:tab3}
\end{table}

\subsection{Validation and discussion}
A validation of the theory is obtained by considering the estimated values for $M$ and $B$
regarding the same magnet N45 obtained from the two test cases 
N45-Al and N45-Cu, as shown in Table~\ref{tab:tab3}. In this case, the relative error is about 10\%. 
Plausible sources of uncertainty in this parameter estimation are the conductivity values for aluminum and copper
used for calibration at room temperature. It is known that the aluminum cylinder is made of an alloy labeled 6082-T6, but 
the exact conductivity corresponding to the correct alloy was not known for any of the two cylinders.
Considering that there are many sources of uncertainty related to this experiment (parameter uncertainties, cylinder edge effects, tilted magnets, etc),  
the relative error of 10\% mentioned above is therefore considered to be satisfactory under the present circumstances (see also the discussion below).

According to the data sheets of the manufacturer the remanence and the coercivity of both the
N45 and the N42 magnets are about $1.3\unit{T}$ and $1152\unit{kAm^{-1}}$, respectively 
(which are related to, but not exactly the same as the present definition of $B$ and $M$).
It should be noted that the manufacturer specify these values based on test objects with different sizes and hence that these values are probably 
subjected to dimensional effects that may depend on the geometry of the test object.

It is also interesting to observe the relatively small induced (displacement) current in the cylinders $I_{\rm ind}\approx 20$-$60\unit{A}$
in comparison to the equivalent current of the magnets $Mh\approx 18$-$20\unit{kA}$,
and which justifies the quasi-static approximation as discussed in section \ref{sect:quasi-static} above.
The characteristic recession velocities for the aluminum cylinder and for the copper cylinder are
$v_0=422\unit{cms^{-1}}$ and $v_0=267\unit{cms^{-1}}$, respectively, 
and which should be compared to the measured velocities displaced in Table~\ref{tab:tab3}.

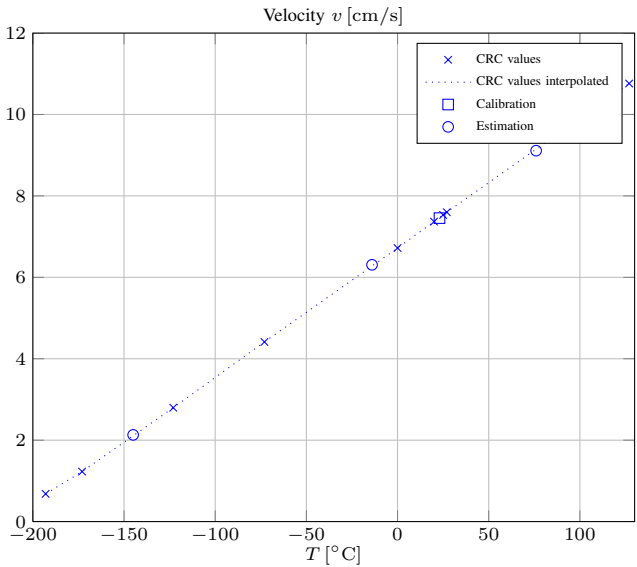
\begin{figure}[htb]
\begin{center}
%\begin{picture}(200,180)
\scriptsize
% This file was created by matlab2tikz v0.4.7 running on MATLAB 7.10.
% Copyright (c) 2008--2014, Nico Schlömer <nico.schloemer@gmail.com>
% All rights reserved.
% Minimal pgfplots version: 1.3
% 
% The latest updates can be retrieved from
%   http://www.mathworks.com/matlabcentral/fileexchange/22022-matlab2tikz
% where you can also make suggestions and rate matlab2tikz.
% 
\begin{tikzpicture}

\begin{axis}[%
width=8cm,
height=6.5cm,
scale only axis,
xmin=-200,
xmax=130,
xlabel style={yshift=0.3cm},
xlabel={$T$\unit{[\degree C]}},
xmajorgrids,
ymin=0,
ymax=12,
ymajorgrids,
title style={yshift=-0.2cm},
title={Velocity $v$\unit{[cm/s]}},
legend style={draw=black,fill=white,legend cell align=left}
]
\addplot [color=blue,only marks,mark=x,mark options={solid}]
  table[row sep=crcr]{-193	0.6812262348702\\
-173	1.22898773801073\\
-123	2.79719833583437\\
-73	4.41267769281228\\
0	6.72050534563785\\
20	7.36836539757563\\
25	7.53241579699336\\
27	7.59914816285819\\
127	10.7605939957048\\
};
\addlegendentry{\tiny CRC values};

\addplot [color=blue,dotted]
  table[row sep=crcr]{-193	0.6812262348702\\
-173	1.22898773801073\\
-123	2.79719833583437\\
-73	4.41267769281228\\
0	6.72050534563785\\
20	7.36836539757563\\
25	7.53241579699336\\
27	7.59914816285819\\
127	10.7605939957048\\
};
\addlegendentry{\tiny CRC values interpolated};

\addplot [color=blue,only marks,mark=square,mark options={solid}]
  table[row sep=crcr]{23	7.45454545454545\\
};
\addlegendentry{\tiny Calibration};

\addplot [color=blue,only marks,mark=o,mark options={solid}]
  table[row sep=crcr]{-14	6.30769230769231\\
};
\addlegendentry{\tiny Estimation};

\addplot [color=blue,only marks,mark=o,mark options={solid},forget plot]
  table[row sep=crcr]{-145	2.12987012987013\\
};
\addplot [color=blue,only marks,mark=o,mark options={solid},forget plot]
  table[row sep=crcr]{76	9.11111111111111\\
};
\end{axis}
\end{tikzpicture}%
%\end{picture}
\end{center}
\vspace{-5mm}
\caption{Temperature estimation in the N45-Al test case: The plot shows the velocity $v$ of the fall vs. cylinder temperature $T$.
The magnetization $M$ is calibrated based on the measured timing data at $T=+23\degree \unit{C}$,
and the corresponding point is indicated with a square in this plot. 
The resistivity data for aluminum at different temperatures from \cite{Lide2008} (the CRC handbook) are incorporated into the model and
indicated with an ``x'', and the corresponding linearly interpolated values are indicated by the dotted line.
The experimental data are then incorporated into the model (inversely mapped) and the corresponding temperature estimations are indicated with an ``o''.}
\label{fig:matfig11a}
\end{figure}

It is illustrative to demonstrate an application of temperature (or resistivity/conductivity) measurements, 
and the fact that the resistivity of conductors is strongly dependent on the temperature.
Hence, the temperature of the measurement cylinders in the four different temperature scenarios described above 
are estimated based on the calibrated magnetization $M$ given in Table~\ref{tab:tab3}. 
The timing data of Table~\ref{tab:tab2} is used to determine the velocity $v$ of the fall and the corresponding
conductivity $\sigma$ of the cylinder is then obtained from \eqref{eq:powerrel}. The
corresponding resistivity $\varrho=1/\sigma$ is then inversely mapped to a temperature $T$ according to the
material data given in \cite{Lide2008}. 
The results are summarized in Table~\ref{tab:tab4}, and the mapping $T\mapsto\varrho\mapsto v$ is illustrated
in Figure \ref{fig:matfig11a} for the test case N45-Al.
It is noted that the resistivity $\varrho$ of the cylinder, and hence the velocity $v$ of the fall, 
are almost linear in the temperature $T$. It is observed that the cylinder is very quickly cooled (heated) when it is brought from the
boiling water at $+100\degree\unit{C}$ (boiling nitrogen at $-196\degree\unit{C}$) to room temperature, due to the large temperature difference
and heat exchange.
As expected,  the temperature estimations 
in the two test cases N45-Al and N42-Al are essentially similar, as the same aluminum cylinder was employed in both cases.

\begin{table}[htb]\centerline{
\begin{tabular}{||c||c|c|c|c||}
\cline{1-5} \scriptsize
\scriptsize Test case  & \scriptsize +100\degree C & \scriptsize +23\degree C & \scriptsize -20\degree C & \scriptsize -196\degree C     \\
\scriptsize temperature   & \scriptsize $T_1$\unit{[\degree C]}  &   \scriptsize$T_2$\unit{[\degree C]}    &    \scriptsize $T_3$\unit{[\degree C]}  &  \scriptsize  $T_4$\unit{[\degree C]}     \\
\cline{1-5} \scriptsize N45-Al      & \scriptsize$76$ & \scriptsize$23$ & \scriptsize$-14$  & \scriptsize$-145$    \\
\cline{1-5} \scriptsize N42-Al     & \scriptsize$73$ & \scriptsize$23$ & \scriptsize$-14$ & \scriptsize$-144$     \\
\cline{1-5} \scriptsize N45-Cu     & \scriptsize$54$ & \scriptsize$23$  & \scriptsize$-8$ & \scriptsize$-170$   \\
\cline{1-5} 
\end{tabular}}
\hspace{5mm}
\caption{Estimated cylinder temperatures.}\label{tab:tab4}
\end{table}

It is emphasized that the purpose of these measurements have been for illustration of the theory
rather than for accurate parameter identification.
Hence, an industrial application for accurate parameter identification would incorporate a much more controlled experiment with an elaborate device
for timing measurements, etc. For practical use, further work would also be needed to obtain error estimates
with respect to the measurement imperfections (timing errors, etc) as well as the modeling imperfections 
(uncertainties regarding parameter values, composition of alloys, the protective coating of the neodymium magnets, 
the permeability of the magnet, the quasi-static approximation, etc).
As for example, the conductivity of the aluminum cylinder is quite uncertain in the present experiment.
If instead of using data from \cite{Lide2008}, the value $\sigma=3.55\cdot10^7$\unit{Sm^{-1}} is used for the conductivity of aluminum at
$T=+20\degree \unit{C}$ as in \cite{Serway1998}, then the estimated parameter values corresponding to N45-Al becomes 
$(M=927\unit{kAm^{-1}},B=0.65\unit{T})$, which is a slightly better match in comparison to the test case N45-Cu.
This example demonstrates that the experiment has a high sensitivity with respect to errors in the ``known'' conductivity.
Hence, for an accurate calibration it is very important to obtain correct information about the
conductivity of the cylinder metal for the particular alloy that is used, etc.
In summary, it is suggested that an accurate measurement procedure would incorporate
\begin{itemize}
\item An elaborate electronic measurement device to facilitate
an accurate timing of the falling magnet.
\item  A measurement set-up that avoids edge effects by using motion detectors inside the cylinder.
\item An inner non-magnetic bearing to keep the magnet horizontally aligned through the fall. 
\item An accurate determination of the conductivity of the cylinder metal used for calibration.
\end{itemize}
If carefully controlled, it is anticipated that the procedure described above
can be used to accurately determine the magnetization of a strong neodymium permanent
magnet based on a cylinder with known electrical properties. Once the magnet has been calibrated, the procedure 
can also be used to accurately determine the conductivity of an arbitrary metal cylinder.

\subsection{Predicted performance}
\begin{figure}[htb]
\begin{center}
%\begin{picture}(200,180)
\scriptsize
% This file was created by matlab2tikz v0.4.7 running on MATLAB 7.10.
% Copyright (c) 2008--2014, Nico Schlömer <nico.schloemer@gmail.com>
% All rights reserved.
% Minimal pgfplots version: 1.3
% 
% The latest updates can be retrieved from
%   http://www.mathworks.com/matlabcentral/fileexchange/22022-matlab2tikz
% where you can also make suggestions and rate matlab2tikz.
% 
\begin{tikzpicture}

\begin{axis}[%
width=8cm,
height=5cm,
scale only axis,
xmin=0,
xmax=30,
xlabel style={yshift=0.3cm},
xlabel={$d$\unit{[mm]}},
xmajorgrids,
ymin=0,
ymax=10,
ymajorgrids,
title style={yshift=-0.2cm},
title={Velocity $v$\unit{[cm/s]}},
legend style={draw=black,fill=white,legend cell align=left}
]
\addplot [color=blue,solid]
  table[row sep=crcr]{2	20.6566506809798\\
2.57142857142857	16.9163891425029\\
3.14285714285714	14.5470804702132\\
3.71428571428571	12.9156024679149\\
4.28571428571428	11.7264999797094\\
4.85714285714286	10.823360632623\\
5.42857142857143	10.1156364987477\\
6	9.54728130725197\\
6.57142857142857	9.08175042323859\\
7.14285714285714	8.69420055163322\\
7.71428571428571	8.36715604855733\\
8.28571428571428	8.08796835584822\\
8.85714285714286	7.84725879600723\\
9.42857142857143	7.63792756859399\\
10	7.45450248439723\\
10.5714285714286	7.29269889976079\\
11.1428571428571	7.14911504386213\\
11.7142857142857	7.02101651292977\\
12.2857142857143	6.90618090425957\\
12.8571428571429	6.80278388257441\\
13.4285714285714	6.70931433899852\\
14	6.62451033166701\\
14.5714285714286	6.54731010396287\\
15.1428571428571	6.47681419805964\\
15.7142857142857	6.41225583974174\\
16.2857142857143	6.35297756304452\\
16.8571428571429	6.29841259406233\\
17.4285714285714	6.24806990157318\\
18	6.2015220995066\\
18.5714285714286	6.15839558685991\\
19.1428571428571	6.11836245737234\\
19.7142857142857	6.08113381970526\\
20.2857142857143	6.04645424983136\\
20.8571428571429	6.01409715833127\\
21.4285714285714	5.9838609016562\\
22	5.9555655019392\\
22.5714285714286	5.92904986736894\\
23.1428571428571	5.90416942647658\\
23.7142857142857	5.88079410639504\\
24.2857142857143	5.85880659832206\\
24.8571428571429	5.83810086386418\\
25.4285714285714	5.81858084427252\\
26	5.80015934126586\\
26.5714285714286	5.78275704352792\\
27.1428571428571	5.76630167733469\\
27.7142857142857	5.75072726332607\\
28.2857142857143	5.73597346434702\\
28.8571428571429	5.72198501167572\\
29.4285714285714	5.70871119892992\\
30	5.69610543457775\\
};
\addlegendentry{Al, $\rho_1=20\unit{mm}$};

\addplot [color=blue,dash pattern=on 1pt off 3pt on 3pt off 3pt]
  table[row sep=crcr]{2	13.0774598013778\\
2.57142857142857	10.7095483392786\\
3.14285714285714	9.20956949965702\\
3.71428571428571	8.17670176512478\\
4.28571428571428	7.42389627746917\\
4.85714285714286	6.85213037558265\\
5.42857142857143	6.40407933119234\\
6	6.04426097126173\\
6.57142857142857	5.74953935757871\\
7.14285714285714	5.50418651963703\\
7.71428571428571	5.29713885211876\\
8.28571428571428	5.12038871557293\\
8.85714285714286	4.96799858993042\\
9.42857142857143	4.83547368287032\\
10	4.71934962180189\\
10.5714285714286	4.6169138539477\\
11.1428571428571	4.52601276196329\\
11.7142857142857	4.44491522999864\\
12.2857142857143	4.37221428349262\\
12.8571428571429	4.30675495925111\\
13.4285714285714	4.24758059368505\\
14	4.19389226763446\\
14.5714285714286	4.14501779664445\\
15.1428571428571	4.1003877455365\\
15.7142857142857	4.05951667941916\\
16.2857142857143	4.0219883650484\\
16.8571428571429	3.98744398515601\\
17.4285714285714	3.95557267419241\\
18	3.92610385953464\\
18.5714285714286	3.89880101919421\\
19.1428571428571	3.87345655993591\\
19.7142857142857	3.84988758836979\\
20.2857142857143	3.82793239883016\\
20.8571428571429	3.80744754047064\\
21.4285714285714	3.78830535535462\\
22	3.7703919018099\\
22.5714285714286	3.75360519468318\\
23.1428571428571	3.73785370763741\\
23.7142857142857	3.72305509321382\\
24.2857142857143	3.70913508471885\\
24.8571428571429	3.69602655060978\\
25.4285714285714	3.68366867732846\\
26	3.67200626076485\\
26.5714285714286	3.66098908994494\\
27.1428571428571	3.65057140930385\\
27.7142857142857	3.64071144815752\\
28.2857142857143	3.63137100782933\\
28.8571428571429	3.6225150984025\\
29.4285714285714	3.61411161831876\\
30	3.60613107107869\\
};
\addlegendentry{Cu, $\rho_1=20\unit{mm}$};

\addplot [color=blue,dotted]
  table[row sep=crcr]{1	15.1697875594843\\
2	8.5634425426441\\
3	6.37388557567167\\
4	5.29007037394094\\
5	4.64878858245618\\
6	4.22856464101314\\
7	3.9343162570215\\
8	3.71844242537063\\
9	3.55448444667067\\
10	3.42657350322433\\
11	3.32462566580548\\
12	3.24193506922377\\
13	3.17387569717394\\
14	3.11715840848069\\
15	3.06938459763743\\
16	3.02876698629024\\
17	2.99394887487682\\
18	2.96388364256351\\
19	2.93775233844209\\
20	2.91490605054448\\
21	2.89482480037662\\
22	2.8770877057321\\
23	2.86135098013475\\
24	2.84733147968494\\
25	2.83479424004532\\
26	2.82354292533906\\
27	2.81341243033103\\
28	2.80426309425135\\
29	2.79597613431873\\
30	2.78845001182385\\
};
\addlegendentry{Al, $\rho_1=16.1\unit{mm}$};

\addplot [color=blue,dashed]
  table[row sep=crcr]{1	9.60383922796878\\
2	5.42142894849464\\
3	4.03524255604677\\
4	3.34909010272894\\
5	2.94310108006837\\
6	2.67706154869477\\
7	2.49077586988323\\
8	2.35410832826006\\
9	2.25030818858093\\
10	2.16932906270066\\
11	2.10478691691447\\
12	2.0524363357266\\
13	2.00934863495557\\
14	1.97344149255688\\
15	1.94319637561988\\
16	1.91748177627741\\
17	1.89543878174471\\
18	1.87640478694706\\
19	1.8598613223392\\
20	1.84539756831072\\
21	1.83268433173094\\
22	1.82145515632758\\
23	1.81149239435619\\
24	1.80261679027483\\
25	1.79467959053559\\
26	1.78755649687864\\
27	1.78114298284785\\
28	1.77535063062094\\
29	1.77010424001927\\
30	1.7653395279119\\
};
\addlegendentry{Cu, $\rho_1=16.1\unit{mm}$};

\addplot [color=blue,only marks,mark=o,mark options={solid},forget plot]
  table[row sep=crcr]{10	7.45\\
};
\addplot [color=blue,only marks,mark=square,mark options={solid},forget plot]
  table[row sep=crcr]{1.4	5.8\\
};
\addplot [color=blue,only marks,mark=diamond,mark options={solid},forget plot]
  table[row sep=crcr]{22	2.8770877057321\\
};
\end{axis}
\end{tikzpicture}%
%\end{picture}
\end{center}
\vspace{-5mm}
\caption{Predicted velocity $v$ of the falling N45 magnet as a function of the cylinder thickness $d$. The plots show
the four combinations corresponding to a cylinder made of aluminum or copper and the inner radius $\rho_1=20\unit{mm}$
or $\rho_1=16.1\unit{mm}$, respectively. The circle and the square indicate the test cases N45-Al and N45-Cu, respectively.
The magnetization is chosen according to the test case N45-Al. The diamond indicates the predicted value for validation.}
\label{fig:matfig2}
\end{figure}
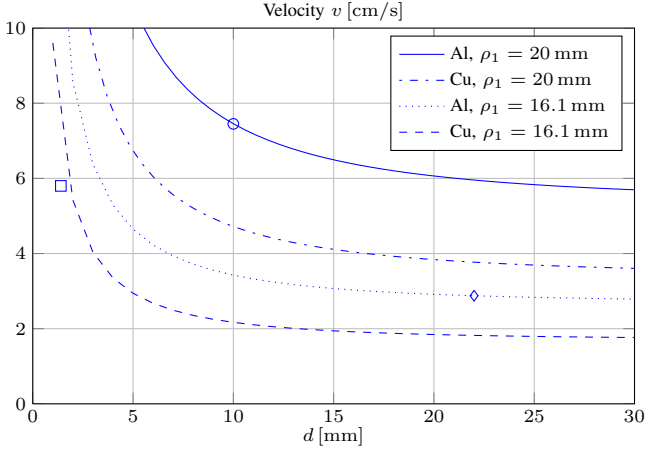

Finally, it is interesting to study the predicted performance of this experiment and how it depends on the cylinder geometry. 
In Figure \ref{fig:matfig2} is shown the predicted velocity of the fall as a function of the cylinder thickness $d=\rho_2-\rho_1$ 
when the cylinder is made of aluminum or copper, and the inner radius is $\rho_1=20\unit{mm}$ or $\rho_1=16.1\unit{mm}$,
respectively. The prediction is performed for the N45 magnet ($a=15\unit{mm}$ and $h=20\unit{mm}$) 
with the cylinders at $T=+23\degree \unit{C}$. 
The results for the N45-Al and N45-Cu test cases are indicated with the circle and the square, respectively.

Simple physical arguments can be used to asses the general limiting behavior of these curves.
As for example, it is reasonable that the (stationary) velocity $v\rightarrow\infty$ when $d\rightarrow 0$,
as there will be no induced current to prevent the acceleration when the cylinder vanishes. 
On the other hand, when $d\rightarrow\infty$ the velocity $v$ must approach a minimum non-zero value, as there must be
induced currents (localized close to the magnet) and corresponding power losses that are associated also with a lossy cylinder of infinite thickness.
Further, as $\rho_1\rightarrow a$, the velocity $v$ must also decrease to a minimum non-zero value, again because there
will always be power losses associated with a lossy cylinder regardless of its geometry. It should be noted, however, that
there will be difficulties to determine the latter limit by numerical computations because of the poor convergence in the
numerical evaluation of the integral in \eqref{eq:Cdef2} as $\rho_1\rightarrow a$. In practice, this is not a problem since in a practical application
$\rho_1-a$ should have a minimum feasible non-zero value.

As an analysis example, it is interesting to note that there is only a slight potential to decrease the velocity of the fall
(by increasing the wall-thickness of the cylinder) for the N45-Al test case (the circle in Figure \ref{fig:matfig2}),
whereas a more significant decrease can be implemented for the N45-Cu test case  (the square in Figure \ref{fig:matfig2}) 
and which is due to the fact that here $\rho_1$ is relatively close to the radius $a$ of the magnet.

Finally, as a validation of the predicted performance, a new aluminum cylinder was manufactured in the same material as before,
with the same length (100\unit{mm}) and with $\rho_1=16.1\unit{mm}$ and $\rho_2=38.1\unit{mm}$ ($d=22\unit{mm}$).
The predicted velocity is $2.88\unit{cms^{-1}}$ and the measured velocity was $2.79\unit{cms^{-1}}$ (3\% error), see also Figure \ref{fig:matfig2}.

\section{Summary}
A rigorous quasi-static electromagnetic analysis has been presented for experiments
and calibration of strong permanent magnets falling inside a metal cylinder. 
The results can be used by teachers and students in electromagnetics who wish to 
obtain a deeper insight into the analysis and experiments regarding this phenomenon.
If the experiment is carefully controlled, the theoretical results can also be employed
with industrial applications such as with an accurate grading or calibration of strong permanent magnets,
or to accurately determine the conductivity of an arbitrary metal cylinder.

%\ack

%The authors acknowledge the comments, suggestions and constructive criticism provided by Patrik Wahlberg.
%The authors acknowledge the comments and suggestions provided by Patrik Wahlberg.
\section*{Acknowledgement}
The authors acknowledge the comments and suggestions provided by Patrik Wahlberg and Yan Levin.

\appendices
\section{Expansion of the free space Green's function}\label{app:ExpGreen}
The free space Green's function for the scalar Helmholtz wave equation satisfies
\begin{equation}\label{eq:Greensreldef1}
\left\{\nabla^2+k^2\right\}G(\vec{r},\vec{r}^\prime)=-\delta(\vec{r}-\vec{r}^\prime),
\end{equation}
where $k$ is the wavenumber of the free space and $\delta(\cdot)$ the three-dimensional Dirac delta function, \cf \cite{Jackson1999}.
When the time-dependence is given by the factor $\eu^{-\iu\omega t}$, 
the Green's function is given explicitly by the outgoing spherical wave
\begin{equation}\label{eq:GreensA0}
G(\vec{r},\vec{r}^\prime)=\frac{\eu^{\iu k|\vec{r}-\vec{r}^\prime|}}{4\pi |\vec{r}-\vec{r}^\prime|},
\end{equation}
where $k=\omega/c$ and where $\omega$ is the angular frequency and $c$ the speed of light in the free space \cite{Jackson1999}.
The Green's function \eqref{eq:GreensA0} can be expanded in cylindrical scalar wave functions as
\begin{multline}\label{eq:GreensA1}
\frac{\eu^{\iu k|\vec{r}-\vec{r}^\prime|}}{4\pi |\vec{r}-\vec{r}^\prime|}
=\frac{\iu}{8\pi}\int_{-\infty}^{\infty}\sum_{m=-\infty}^{\infty}\\
{\rm J}_{m}(\kappa\rho_<){\rm H}_{m}^{(1)}(\kappa\rho_>)\eu^{\iu m(\phi-\phi^\prime)}\eu^{\iu\alpha(z-z^\prime)}\diff\alpha,
\end{multline}
where $\alpha$ is the Fourier variable corresponding to a Fourier transformation along the longitudinal coordinate $z$,
and ${\rm J}_m(\cdot)$ and ${\rm H}_m^{(1)}(\cdot)$ are the regular Bessel functions and 
the Hankel functions of the first kind, both of order $m$, respectively, 
see \eg \cite{Collin1991,Jackson1999}. Here, the
transverse wavenumber is defined by
\begin{equation}\label{eq:kappadef}
\kappa=\sqrt{k^2-\alpha^2},
\end{equation}
where the square root\footnote{If the square root is defined as \eg with the MATLAB software where
$-\pi/2<\arg\sqrt{w}\leq\pi/2$ for $-\pi<\arg w\leq \pi$, then $\kappa$ can be defined here as $\kappa=\iu\sqrt{-k^2+\alpha^2}$
which implies that $0<\arg\kappa\leq \pi$.} $\kappa=\sqrt{w}$ is defined such that $0<\arg w\leq 2\pi$ and
$0<\arg\kappa\leq \pi$ and hence $\Im\kappa\geq 0$. 
The arguments of the cylindrical functions above are furthermore defined by 
$\rho_<=\min\{\rho,\rho^\prime\}$ and $\rho_>=\max\{\rho,\rho^\prime\}$,
and where the primed variables represent the source point and the unprimed variables represent the field point.
It is noted that the expansion in \eqref{eq:GreensA1} involves regular Bessel functions 
${\rm J}_{m}(\kappa\rho)$ for $\rho<\rho^\prime$ and outgoing (radiating) Hankel functions ${\rm H}_{m}^{(1)}(\kappa\rho)$
for $\rho>\rho^\prime$. It is also noted that the combination ${\rm J}_{m}(\kappa\rho_<){\rm H}_{m}^{(1)}(\kappa\rho_>)$
is continuous across the point where $\rho=\rho^\prime$ with its value ${\rm J}_{m}(\kappa\rho^\prime){\rm H}_{m}^{(1)}(\kappa\rho^\prime)$.

To verify that the expansion in \eqref{eq:GreensA1} satisfies \eqref{eq:Greensreldef1}, it is first noted that the cylindrical scalar wave functions
$\psi_m(\kappa\rho)\eu^{\iu m\phi}\eu^{\iu\alpha z}$ satisfy the homogeneous Helmholtz wave equation
\begin{equation}
\left\{\nabla^2+k^2\right\}\psi_m(\kappa\rho)\eu^{\iu m\phi}\eu^{\iu\alpha z}=0,
\end{equation}
where $\psi_m(\kappa\rho)$ is any cylindrical function of order $m$ satisfying
the Bessel differential equation \cite{Olver+etal2010,Arfken+etal2013,Jackson1999}
\begin{equation}\label{eq:Bessel}
\left\{\frac{1}{\rho}\frac{\partial}{\partial\rho}\rho\frac{\partial}{\partial\rho}+\kappa^2-\frac{m^2}{\rho^2} \right\}\psi_m(\kappa\rho)=0.
\end{equation}
Consider now the distributional relationship \eqref{eq:Greensreldef1} in cylindrical coordinates
\begin{multline}\label{eq:Greensreldef2}
\left\{\frac{1}{\rho}\frac{\partial}{\partial\rho}\rho\frac{\partial}{\partial\rho}
+\frac{1}{\rho^2}\frac{\partial^2}{\partial \phi^2}+\frac{\partial^2}{\partial z^2}
+k^2\right\}G(\rho,\phi,z,\rho^\prime,\phi^\prime,z^\prime) \\
=-\frac{1}{\rho}\delta(\rho-\rho^\prime)\delta(\phi-\phi^\prime)\delta(z-z^\prime).
\end{multline}
Take the Fourier transform of \eqref{eq:Greensreldef2} to yield
\begin{multline}\label{eq:Greensreldef3}
\left\{\frac{1}{\rho}\frac{\partial}{\partial\rho}\rho\frac{\partial}{\partial\rho}
-\frac{m^2}{\rho^2}-\alpha^2
+k^2\right\}G_m(\rho,\alpha,\rho^\prime,\phi^\prime,z^\prime) \\
=-\frac{1}{\rho}\delta(\rho-\rho^\prime)\frac{1}{2\pi}\eu^{-\iu m\phi^\prime}\eu^{-\iu\alpha z^\prime},
\end{multline}
where the following relationships
 have been used $\frac{\partial}{\partial\phi}\leftrightarrow \iu m$, $\frac{\partial}{\partial z}\leftrightarrow \iu \alpha$,
$\sum_m\eu^{\iu m(\phi-\phi^\prime)}=2\pi\delta(\phi-\phi^\prime)$ and $\int \eu^{\iu\alpha(z-z^\prime)}\diff\alpha=2\pi\delta(z-z^\prime)$.

From \eqref{eq:GreensA1}, it follows that the conjecture to be proven is
\begin{equation}\label{eq:Greensreldef4}
G_m(\rho,\alpha,\rho^\prime,\phi^\prime,z^\prime) \\
=\frac{\iu}{4}{\rm J}_{m}(\kappa\rho_<){\rm H}_{m}^{(1)}(\kappa\rho_>)\eu^{-\iu m\phi^\prime}\eu^{-\iu\alpha z^\prime},
\end{equation}
and it will now be verified that \eqref{eq:Greensreldef4} satisfies the distributional relationship \eqref{eq:Greensreldef3}.
Hence, by inserting \eqref{eq:Greensreldef4} into \eqref{eq:Greensreldef3}, it follows that
\begin{multline}\label{eq:Greensreldef5}
\left\{\frac{1}{\rho}\frac{\partial}{\partial\rho}\rho\frac{\partial}{\partial\rho}
-\frac{m^2}{\rho^2}-\alpha^2
+k^2\right\}{\rm J}_{m}(\kappa\rho_<){\rm H}_{m}^{(1)}(\kappa\rho_>) \\
=\iu\frac{2}{\pi}\frac{1}{\rho}\delta(\rho-\rho^\prime).
\end{multline}
It is noted that ${\rm J}_{m}(\kappa\rho_<){\rm H}_{m}^{(1)}(\kappa\rho_>)$ is a continuous function across $\rho=\rho^\prime$,
and with a discontinuous derivative.
Hence, by applying to \eqref{eq:Greensreldef5} the integration $\int_{\rho^\prime-}^{\rho^\prime+}\{\cdot\}\rho\diff\rho$
it follows that
\begin{equation}\label{eq:Greensreldef6}
\left[\rho\frac{\partial}{\partial\rho} {\rm J}_{m}(\kappa\rho_<){\rm H}_{m}^{(1)}(\kappa\rho_>) \right]_{\rho^\prime-}^{\rho^\prime+}
=\iu\frac{2}{\pi},
\end{equation}
where $\rho^\prime+$ and $\rho^\prime-$ means taking a limit towards $\rho^\prime$ from above and from below, respectively.
It follows that
\begin{equation}\label{eq:Greensreldef7}
\kappa\rho^\prime\left[  {\rm J}_{m}(\kappa\rho^\prime){\rm H}_{m}^{(1)\prime}(\kappa\rho^\prime)
- {\rm J}_{m}^\prime(\kappa\rho^\prime){\rm H}_{m}^{(1)}(\kappa\rho^\prime)\right] 
=\iu\frac{2}{\pi},
\end{equation}
where ${\rm J}_{m}^\prime(\cdot)$ and ${\rm H}_{m}^{(1)\prime}(\cdot)$ denote a differentiation with respect to the argument.
The validity of \eqref{eq:Greensreldef7}, and hence of \eqref{eq:Greensreldef3}, is finally verified
by using the Wronskian relation ${\cal W}({\rm J}_{m}(\zeta),{\rm H}_{m}^{(1)}(\zeta))=2\iu/(\pi\zeta)$, \cf \cite{Olver+etal2010,Arfken+etal2013}.

The analysis and results given above are also valid when $k=0$, and the expansion used in \eqref{eq:Greens1} is 
hence given by \eqref{eq:GreensA1} with $\kappa=\iu|\alpha|$ and where $|\cdot|$ denotes the absolute value.

%\bibliographystyle{teorel} % For arxiv
%\bibliography{total}
%\input{PJmagnet.bbl}

\begin{IEEEbiographynophoto}{Sven~Nordebo}
received the M.S.\ degree in electrical engineering from the Royal Institute of Technology, Stockholm, Sweden, in 1989, 
and the Ph.D.\ degree in signal processing from Lule{\aa} University of Technology, Lule{\aa}, Sweden, in 1995. 
Since 2002 he is a Professor of Signal Processing at the 
Department of Physics and Electrical Engineering, Linn\ae us University. 
His research interests are in statistical signal processing, electromagnetic wave propagation, inverse problems and imaging.
\end{IEEEbiographynophoto}%\vspace{-1.4cm}

\begin{IEEEbiographynophoto}{Alexander~Gustafsson}
received the M.S.\ degree in physics from the Linn\ae us University, V\"{a}xj\"{o}, Sweden, in 2011. 
Since 2013 he is a Ph.D student in physics at the 
Department of Physics and Electrical Engineering, Linn\ae us University. 
His research interests are in molecular dynamics on surfaces.
\end{IEEEbiographynophoto}%\vspace{-1.4cm}

\end{document}